\documentclass[useAMS,usenatbib]{mn2e}

\pdfoutput=1

\newcommand{\thisstar}{KIC\,8462852}	
\newcommand{\kepler}{{\it Kepler}}	
\newcommand{\rsun}{R$_{\odot}$}			% R_sun
\newcommand{\msun}{M$_{\odot}$}			% M_sun
\newcommand{\lsun}{L$_{\odot}$}			% L_sun
\newcommand{\mearth}{M$_{\oplus}$}		% M_earth

\newcommand{\kms}{km~s$^{-1}$}			% km/s
\newcommand{\perone}{\mbox{$^{-1}$}}
\newcommand{\degs}{\mbox{$^{\circ}$}}

\newcommand{\phn}{\phantom{0}}		% phantom number
\def\apjl{{ApJ}}                % Astrophysical Journal, Letters
\def\apjs{{ApJS}}               % Astrophysical Journal, Supplement
\def\aap{{A\&A}}                % Astronomy and Astrophysics
\usepackage{times}
\usepackage{epsfig}
\usepackage{graphicx}              % For powerful manipulation of figures.
\usepackage{amsmath,amsfonts,amsopn,amssymb} % Some nice math packages.
\usepackage{afterpage}
\usepackage{deluxetable}
\usepackage{natbib}                % To format bibliographies.
\usepackage{times}
\usepackage{array}
\usepackage{caption}
\usepackage{subcaption}
\usepackage[usenames, dvipsnames]{color}

\usepackage[backref,breaklinks,colorlinks,citecolor=blue]{hyperref}        
\usepackage[all]{hypcap}

\definecolor{Blue}{rgb}{0.3,0.3,0.9}
\usepackage{color}

\title[KIC\,8462852 -- Where's the flux?]
{Planet Hunters X. \\
KIC\,8462852 -- Where's the flux?
\thanks{Based on observations obtained with the Nordic Optical Telescope, operated on the island of La Palma jointly by Denmark, Finland, Iceland, Norway, and Sweden, in the Spanish Observatorio del Roque de los Muchachos of the Instituto de Astrofisica de Canarias.}\thanks{The data presented herein were obtained at the W.M. Keck Observatory, which is operated as a scientific partnership among the California Institute of Technology, the University of California, and the National Aeronautics and Space Administration. The Observatory was made possible by the generous financial support of the W.M. Keck Foundation.}}
\author[ ]{\parbox{\textwidth}
{T. S. Boyajian$^{1}$,
D. M. LaCourse$^{2}$,
S. A. Rappaport$^{3}$,\\
D. Fabrycky$^{4}$,
D. A. Fischer$^{1}$,
D. Gandolfi$^{5,6}$,
G. M. Kennedy$^{7}$,
H. Korhonen$^{8,9}$,	
M. C. Liu$^{10}$,		
A. Moor$^{11}$,		
K. Olah$^{11}$,			
K. Vida$^{11}$,			
M. C. Wyatt$^{7}$,
W. M. J. Best$^{10}$,	
J. Brewer$^{1}$,
F. Ciesla$^{12}$,		
B. Cs\'{a}k$^{13}$,		
H. J. Deeg$^{14,15}$,	
T. J. Dupuy$^{16}$,		
G. Handler$^{17}$,		
K. Heng$^{18}$,				
S. B. Howell$^{19}$,	
S. T. Ishikawa$^{20}$,	
J. Kov\'{a}cs$^{13}$,		
T. Kozakis$^{21}$,		
L. Kriskovics$^{11}$,	
J. Lehtinen$^{22}$,		
C. Lintott$^{23}$,		
S. Lynn$^{24}$,			
D. Nespral$^{14,15}$,	
S. Nikbakhsh$^{22,25}$,	
K. Schawinski$^{26}$,	
J. R. Schmitt$^{1}$,
A. M. Smith$^{27}$,		
Gy. Szabo$^{11,13,28}$,	
R. Szabo$^{11}$,		
J. Viuho$^{22}$,		
J. Wang$^{1,29}$,		
A. Weiksnar$^{20}$,	
M. Bosch$^{2}$,
J. L. Connors$^{2}$,
S. Goodman$^{2}$,
G. Green$^{2}$,
A. J. Hoekstra$^{2}$,
T. Jebson$^{2}$,
K. J. Jek$^{2}$,
M. R. Omohundro$^{2}$,
H. M. Schwengeler$^{2}$,
A. Szewczyk$^{2}$
}
\vspace{0.3cm}\\
$^{1}$Department of Astronomy, Yale University, New Haven, CT 06511, USA\\
$^{2}$Amateur Astronomer\\
$^{3}$Department of Physics, and Kavli Institute for Astrophysics and Space Research, Massachusetts Institute of Technology, Cambridge, MA 02139, USA\\
$^{4}$Department of Astronomy and Astrophysics, University of Chicago, 5640 South Ellis Avenue, Chicago, IL 60637, USA\\
$^{5}$Dipartimento di Fisica, Universit\'a di Torino, via P. Giuria 1, I-10125, Torino, Italy\label{Torino}\\
$^{6}$Landessternwarte K\"onigstuhl, Zentrum f\"ur Astronomie der Universit\"at Heidelberg, K\"onigstuhl 12, D-69117 Heidelberg, Germany\label{LSW}\\
$^{7}$Institute of Astronomy, University of Cambridge, Madingley Road, Cambridge CB3 0HA, UK\\
$^{8}$Finnish Centre for Astronomy with ESO (FINCA), University of Turku, V\"{a}is\"{a}l\"{a}ntie 20, FI-21500 Piikki\"{o}, Finland\\
$^{9}$Centre for Star and Planet Formation, Niels Bohr Institute, University of Copenhagen, {\O}ster Voldgade 5-7, DK-1350, K{\o}benhavn K, Denmark\\
$^{10}$Institute for Astronomy, University of Hawaii, 2680 Woodlawn Drive, Honolulu HI 96822, USA\\
$^{11}$Konkoly Observatory, Research Centre of Astronomy and Earth Sciences, Hungarian Academy of Sciences, H-1121 Budapest, Konkoly Th. M. \'{u}t 15 --17, Hungary\\
$^{12}$Department of the Geophysical Sciences, The University of Chicago, 5734 South Ellis Avenue, Chicago, IL 60637\\
$^{13}$ELTE Gothard Astrophysical Observatory, H-9704 Szombathely, Szent Imre herceg ut 112, Hungary\\
$^{14}$Instituto de Astrof\'{i}sica de Canarias, C. V\'{i}a L\'{a}ctea S/N, E-38205 La Laguna, Tenerife, Spain\\
$^{15}$Departamento de Astrof\'{i}sica, Universidad de La Laguna, E-38200 La Laguna, Tenerife, Spain\\
$^{16}$The University of Texas at Austin, Department of Astronomy, 2515 Speedway C1400, Austin, TX 78712, USA \\
$^{17}$Copernicus Astronomical Center, Bartycka 18, 00-716 Warsaw, Poland\\
$^{18}$University of Bern, Center for Space and Habitability, Sidlerstrasse 5, CH-3012, Bern, Switzerland\\
$^{19}$NASA Ames Research Center, Moffett Field, CA 94035, USA\\
$^{20}$Adler Planetarium, Department of Citizen Science, 1300 S Lake Shore Dr, Chicago, IL 60605\\
$^{21}$Carl Sagan Institute, Cornell University, Ithaca, NY 14853, USA\\
$^{22}$Department of Physics, PO Box 64, 00014 University of Helsinki, Finland\\
$^{23}$Department of Physics, University of Oxford, Denys Wilkinson Building, Keble Road, Oxford, OX1 3RH, UK\\
$^{24}$CartoDB, 247 Centre Street, New York, NY 10013, USA\\
$^{25}$Finnish Meteorological Institute, Post Office Box 503, FI-00101 Helsinki, Finland\\
$^{26}$Institute for Astronomy, Department of Physics, ETH Zurich, Wolfgang-Pauli-Strasse 27, CH-8093 Zurich, Switzerland\\
$^{27}$GitHub Inc, 88 Colin P Kelly Jr St, San Francisco, CA 94107 USA\\
$^{28}$Gothard-Lend\"{u}let Research Team, H-9704 Szombathely, Szent Imre herceg \'{u}t 112, Hungary\\
$^{29}$California Institute of Technology, Pasadena, CA 91109, USA
}

\begin{document}
\maketitle
\label{firstpage}
%%%%%%%%%%%%%%%%%%%%%%%%%%%%%%%%%%%%%%%%%%%%%%%%%%%%%%%%%%%%%%%%
%
\clearpage
\begin{abstract}
\noindent
Over the duration of the \kepler\ mission, \thisstar\ was observed to undergo irregularly shaped, aperiodic dips in flux of up to $\sim 20$\%.  The dipping activity can last for between 5 and 80 days.  We characterize the object with high-resolution spectroscopy, spectral energy distribution fitting, radial velocity measurements, high-resolution imaging, and Fourier analyses of the \kepler\ light curve.  We determine that \thisstar\ is a typical main-sequence F3~V star that exhibits no significant IR excess, and has no very close interacting companions.  In this paper, we describe various scenarios to explain the dipping events observed in the \kepler\ light curve.  We confirm that the dipping signals in the data are not caused by any instrumental or data processing artifact, and thus are astrophysical in origin.  We construct scenario-independent constraints on the size and location of a body in the system that is needed to reproduce the observations.  We deliberate over several assorted stellar and circumstellar astrophysical scenarios, most of which have problems explaining the data in hand.  By considering the observational constraints on dust clumps in orbit around a normal main-sequence star, we conclude that the scenario most consistent with the data in hand is the passage of a family of exocomet or planetesimal fragments, all of which are associated with a single previous break-up event, possibly caused by tidal disruption or thermal processing.  The minimum total mass associated with these fragments likely exceeds $10^{-6}$~\mearth, corresponding to an original rocky body of $>100$~km in diameter.  We discuss the necessity of future observations to help interpret the system.     

\end{abstract}
%%%%%%%%%%%%%%%%%%%%%%%%%%%%%%%%%%%%%%%%%%%%%%%%%%%%%%%%%%%%%%%%

\begin{keywords}
stars: individual (\thisstar), stars: peculiar, stars: activity, comets: general, planets and satellites: dynamical evolution and stability
\end{keywords}

%%%%%%%%%%%%%%%%%%%%%%%%%%%%%%%%%%%%%%%%%%%%%%%%%%%%%%%%%%%%%%%%

\section{Introduction}\label{sec:introduction}

For over four years, NASA's \kepler\ mission measured the brightness of objects within a $\sim 100$~square-degree patch of sky in the direction of the constellations Cygnus and Lyrae.  The program's targets were primarily selected to address the \kepler\ mission goals of discovering Earth-like planets orbiting other stars.  \kepler\ targeted over $>150,000$ stars, primarily with a 30-minute observing cadence, leading to over 2.5-billion data points per year ($>10$~billion data points over the nominal mission lifetime). 

The \kepler\ mission's data processing and identification of transiting planet candidates was done in an automated manner through sophisticated computer algorithms (e.g., \citealt{jen10}). Complementary to this analysis, the Zooniverse citizen science network provided the means to crowd source the review of light curves with the Planet Hunters project\footnote{www.planethunters.org} \citep[e.g.,][]{fis12}. In this framework, Planet Hunter volunteers view 30 day segments of light curves in the `Classify' web interface.  A volunteer's main task is to identify signals of transiting planets by harnessing the human eye's unique ability for pattern recognition.  This process has shown to have a detection efficiency to identify planetary transits $> 85$\% using the first Quarter of \kepler\ data \citep{sch12}.  The Planet Hunters project has now discovered almost a hundred exoplanet candidates, including several confirmed systems \citep{fis12, lin13,sch13,wan13, sch14}.     

Because Planet Hunter volunteers look at every light curve by eye, serendipitous discoveries are inevitable, especially in rich data sets such as that which \kepler\ has provided.  As such, a key aspect of the Planet Hunters project is the `Talk' interface.  `Talk' is a community discussion board/site where volunteers can discuss light curves and present further analysis on objects viewed in the main `Classify' interface.  In a handful of cases, such as the discovery of the unusual cataclysmic variable, KIC\,9406652 \citep{gie13}, the default aperture mask used to generate the \kepler\ light curve was not perfectly centered on the object of interest. Because of this, interesting events in the \kepler\ light curve would appear to come and go as a result of the shifting orientation of the aperture mask when the spacecraft underwent a quarterly rotation.  Events such as these are tagged and discussed on `Talk', making it possible to return to the raw data target pixel files (TPF) to extract improved light curves with modified aperture masks, for example.   

This paper presents the discovery of a highly unusual dipping source, \thisstar, from the Planet Hunters project.  In just the first quarter of \kepler\ data, Planet Hunter volunteers identified \thisstar's light curve as a ``bizarre'', ``interesting'', ``giant transit'' (Q1 event depth was 0.5\% with a duration of $~4$~days).  As new \kepler\ data were released in subsequent quarters, discussions continued on `Talk' about \thisstar's light curve peculiarities, particularly ramping up pace in the final observations of the Kepler mission.

In this work we examine the full 4 years of {\em Kepler} observations of \thisstar\ as well as supplemental data provided by additional ground- and space-based observations.  In Section~\ref{sec:data}, we characterize \thisstar\ using \kepler\ photometry, spectroscopic analysis, AO imaging, and spectral energy distribution analysis.  We discover a wide M-dwarf companion to the system and argue that with the data sets we have in-hand, we can exclude the presence of an additional massive gravitationally bound companion nearby.  In Section~\ref{sec:features}, we visit possible explanations for the peculiar observations of \thisstar, including instrumental artifacts, intrinsic/extrinsic variability, and a variety of scenarios invoking light-blocking events. We formulate a variety of  scenario independent constraints in Section~\ref{ss:indep}, and elaborate on specific occultation scenarios in Section~\ref{ss:specific_occ_scen}.  In Section~\ref{sec:discussion}, we conclude by discussing future observations needed to constrain the nature of the object.

%%%%%%%%%%%%%%%%%%%%%%%%%%%%%%%%%%%%%%%%%%%%%%%%%%%%%%%%%%
\section{Data}\label{sec:data}

\thisstar, also known as TYC 3162-665-1 and 2MASS J20061546+4427248, is a $V \sim 12$ mag star in the \kepler\ field of view.  Its light curve was identified serendipitously by the Planet Hunters project, and was deemed an interesting object that was worthy of further investigation.  In the following sections, we characterize the system with data from \kepler\, as well as additional data from various targeted and archived programs.

%%%%%%%%%
\subsection{{\it Kepler} photometry}\label{sec:kepler}

The \kepler\ mission was launched on 2009 March 7, and it started science observations on 2009 May 13.  The nominal mission was finished almost 4 years later, on 2013 May 12, after the failure of the second reaction wheel. \thisstar\ was observed throughout the main \kepler\ mission (divided into Quarters 0 -- 17) under long-cadence (30-minute) observations yielding an ultra-precise, nearly uninterrupted, light curve during this time.  \kepler\ data files provide both the `uncorrected' Simple Aperture Photometry (SAP) and the `corrected' Pre-search Data Conditioning (PDCSAP) fluxes (for details, see \citealt{chr12}).  In this work, our analysis uses the normalized, PDCSAP data.   Note that we have thoroughly validated the data to ensure that any flux variations represent physical events in or near the star (and they do); these processes are described in detail within Section~\ref{sec:instrumental}, and we do not repeat them here.   

%%%%%%%%%%%%%%%%%%%%%%%%%%%%%%%%%%%%%%%%%%%%%%%%%%%%%%%%%%
% photometry montage plot 
\begin{figure*}
    \begin{center}
        \begin{tabular}{cc}
\multicolumn{2}{c}{\includegraphics[angle=0,width=0.8\textwidth]{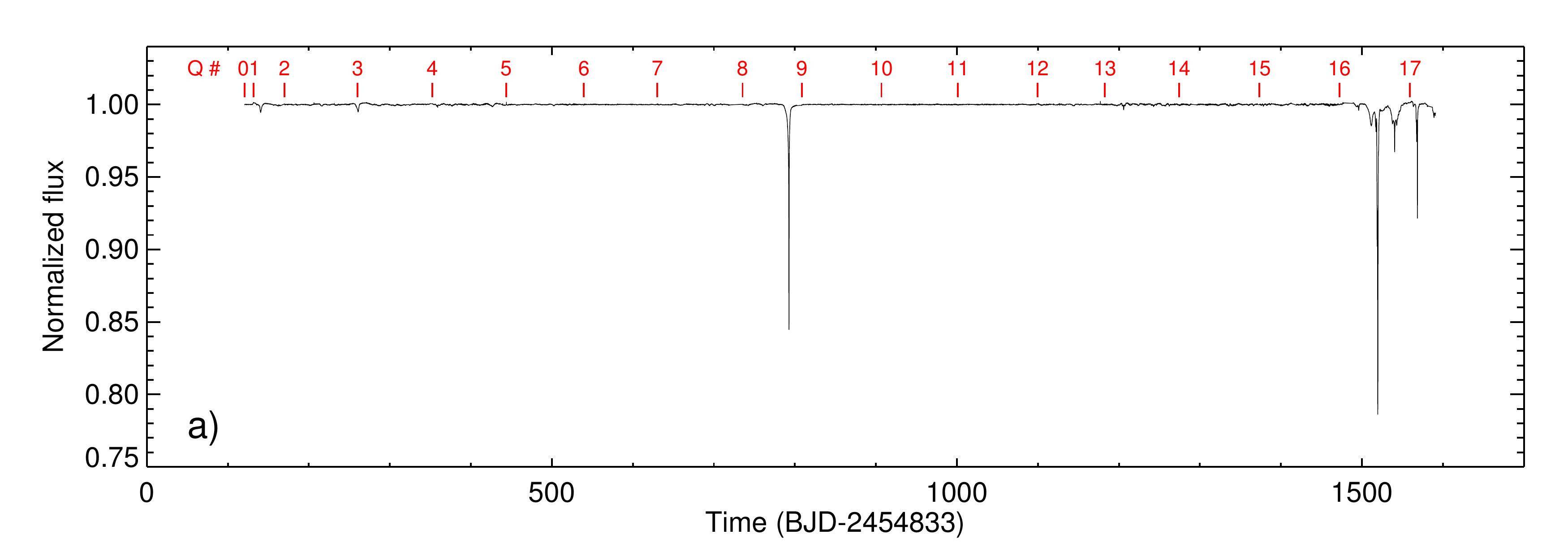}}	\\
\multicolumn{2}{c}{\includegraphics[angle=0,width=0.8\textwidth]{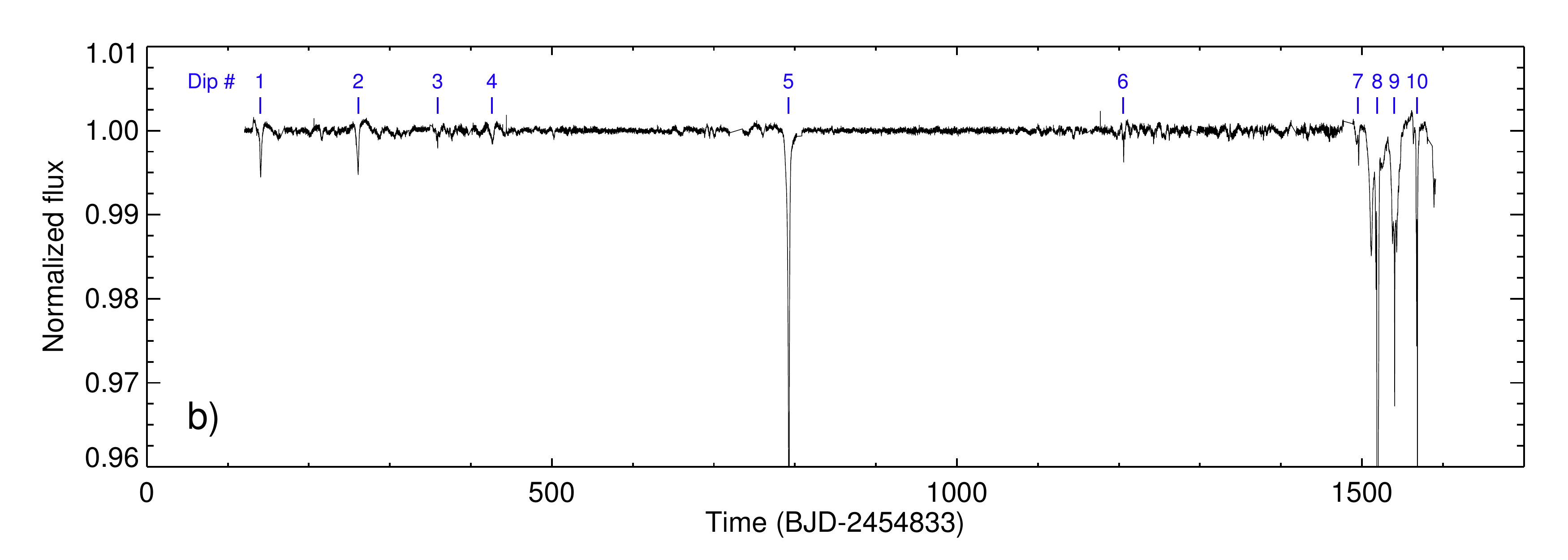}}	\\
            \includegraphics[angle=0,width=0.4\textwidth]{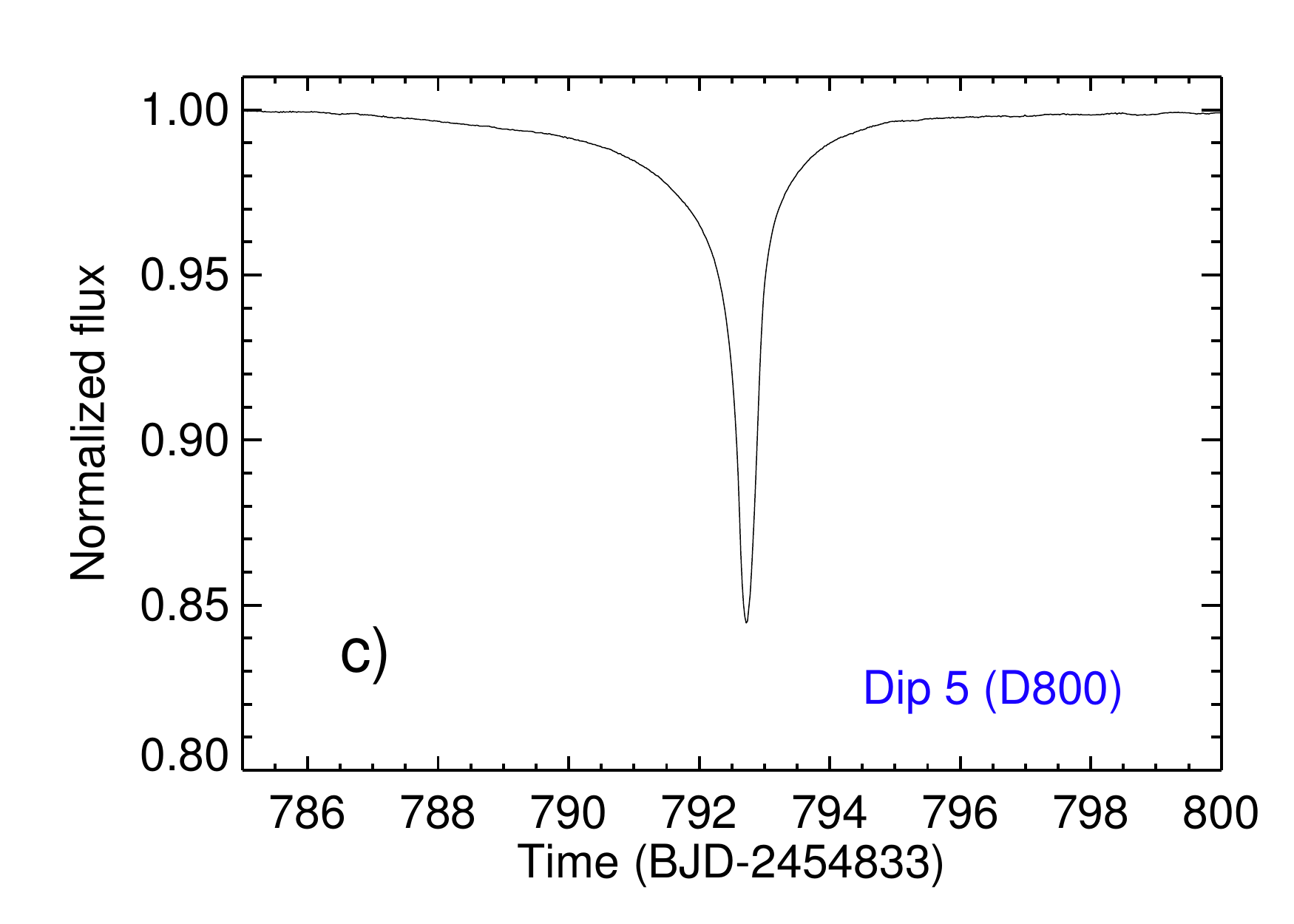}	&
            \includegraphics[angle=0,width=0.4\textwidth]{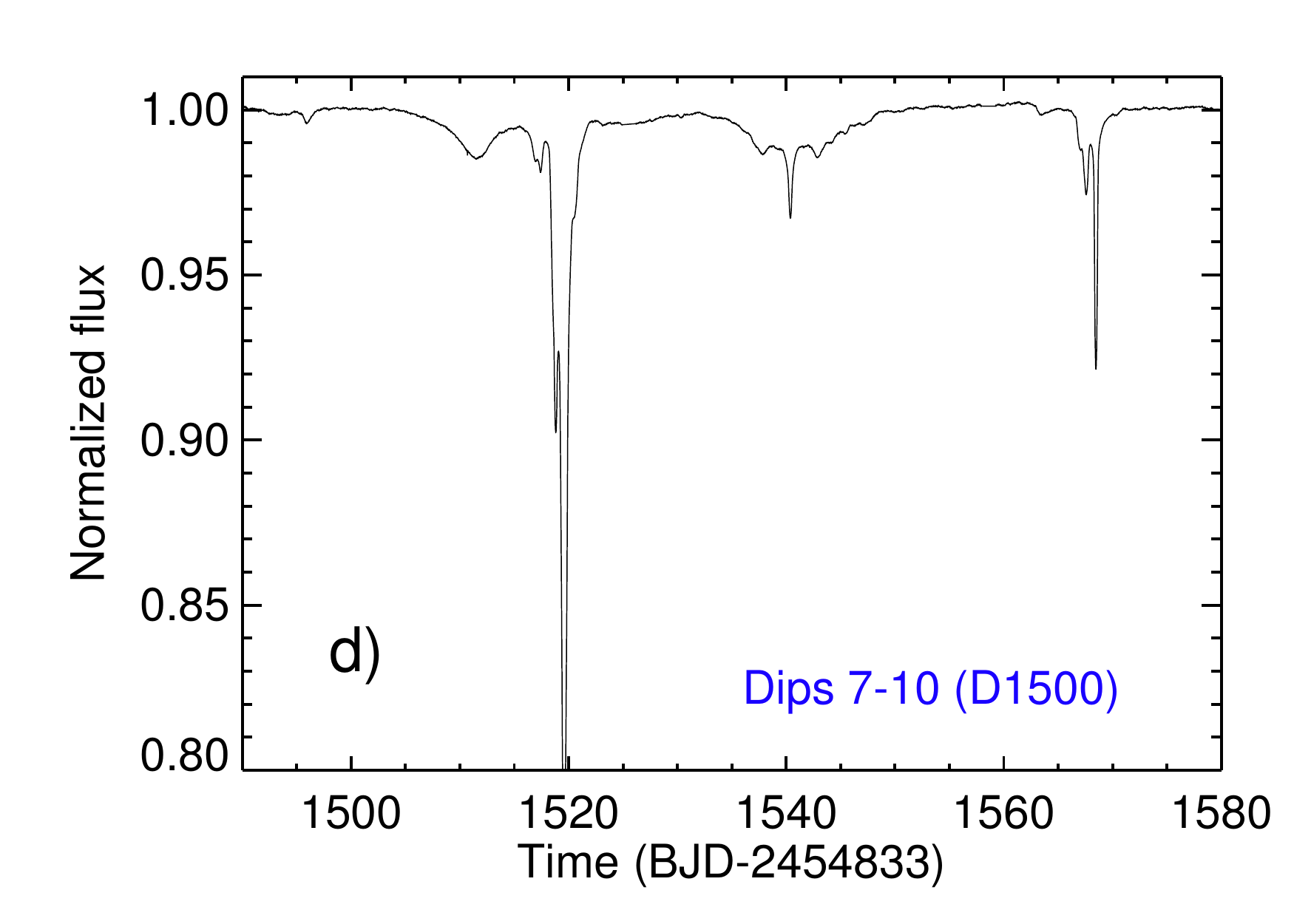}	\\
            \includegraphics[angle=0,width=0.4\textwidth]{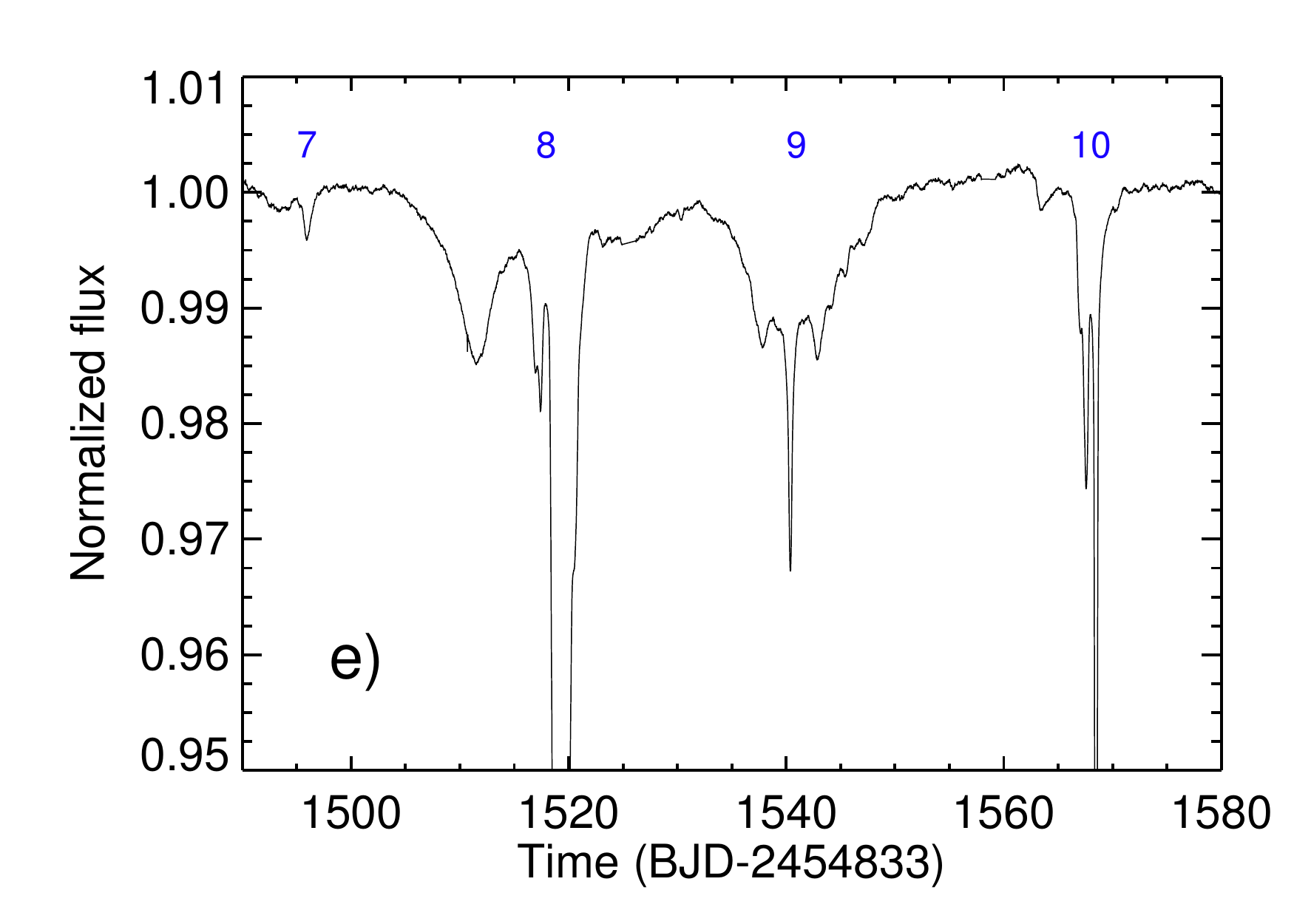}	&
            \includegraphics[angle=0,width=0.4\textwidth]{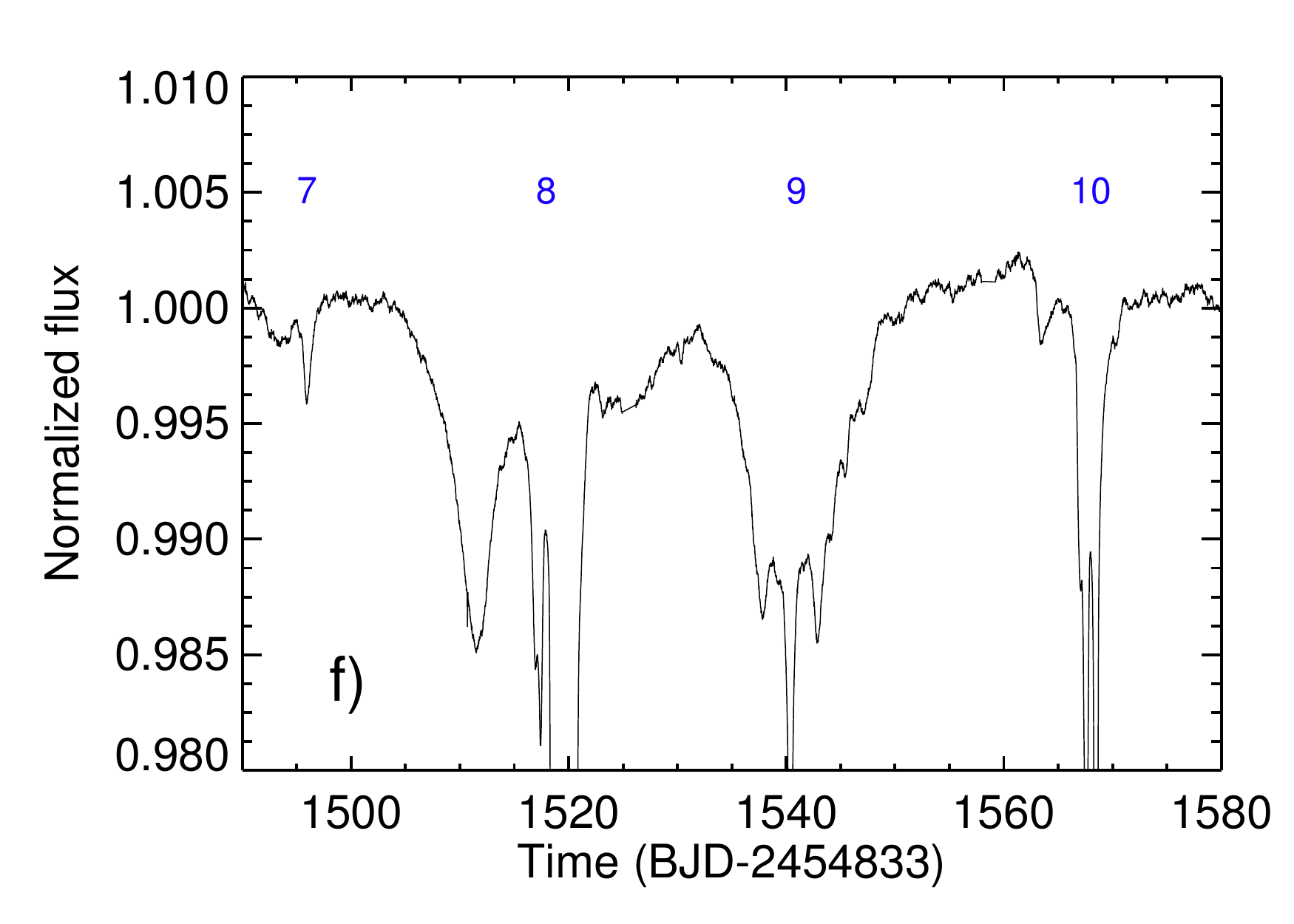}	           
        \end{tabular}
        \end{center}
    \caption{Montage of flux time series for \thisstar\ showing different portions of the 4-year {\em Kepler} observations with different vertical scalings. The top two panels show the entire \kepler\ observation time interval.  The starting time of each \kepler\ quarter Q is marked and labeled in red in the top panel `(a)'.  Dip numbers corresponding to the 10 discrete dips listed in Table~\ref{tab:Dip-Table} are marked and labeled in blue in panel `(b)'.  Panel `(c)' is a blowup of the dip \#~5 near day 793 (D800). The remaining three panels, `(d)', `(e)', and `(f)', explore the dips (labeled in blue) which occur during the 90-day interval from day 1490 to day 1580 (D1500).  Refer to Section~\ref{sec:kepler} for details. }
    \label{fig:kepler}
  \end{figure*} 
%%%%%%%%%%%%%%%%%%%%%%%%%%%%%%%%%%%%%%%%%%%%%%%%%%%%%%%%%%

In Figure~\ref{fig:kepler}, we present a montage of plots capturing much of the interesting flux variations observed in the \kepler~timeseries data.  The top two panels, `(a)' and `(b)', show the flux time series for the entire \kepler\ mission, but with different vertical flux scales.  These show that the flux is relatively constant for most of that time, but is punctuated by a number of substantial dips in flux, including a 15\% drop near day 800, and a whole sequence of dips (with one reaching a depth of 22\%) after day 1500.  Panel `(b)' marks the occurrence of 10 discrete dips (see Table~\ref{tab:Dip-Table}).  For convenience, we hereafter refer to the two main dip structures between day 788 and 795 and between day 1510 and 1570, as events `D800' and `D1500', respectively.  Panel `(c)' is a zoom in on the dip D800.  The remaining three panels are progressively zoomed in around the exotic complex of dips at D1500. 

The D800 dip feature is clean, sharp, and asymmetric in shape. It possesses a gradual dimming lasting almost a week, and transitions back to its nominal brightness in just a couple of days.  The D1500 complex consists of many dips, with variable shape and duration, often occurring concurrently as if several independent occultation events were superimposed upon each other.  The D1500 dips persist for $\sim 100$ days until the \kepler\ mission's end, and only for a small part of this time does it appear `quiescent'.  
There are also other smaller $\sim 0.5$\% dips, including three earlier in the mission around day 140, day 260, and day 359, and another after the D800 event, around day 1205 (dips \#1, 2, 3 and 6, respectively; Figure~\ref{fig:kepler} `(b)', Table~\ref{tab:Dip-Table}).  Several more $0.5-1$\% dips appear in and around the two deep D1500 features, including a $\sim 3$\% dip around day 1540. Two small dips occurring at day 1205 and day 1540 have shapes with a similar distinctive, `triple-dip', symmetric profile, however, they differ in duration by a factor of 3 and in degree of dimming by a factor of 5.  All of the fluctuations in intensity visible on these plots are real, i.e., not due to statistical or instrumental variations (Section~\ref{sec:instrumental}).

\begin{figure}
  \centering  
      \includegraphics[width=84mm]
      {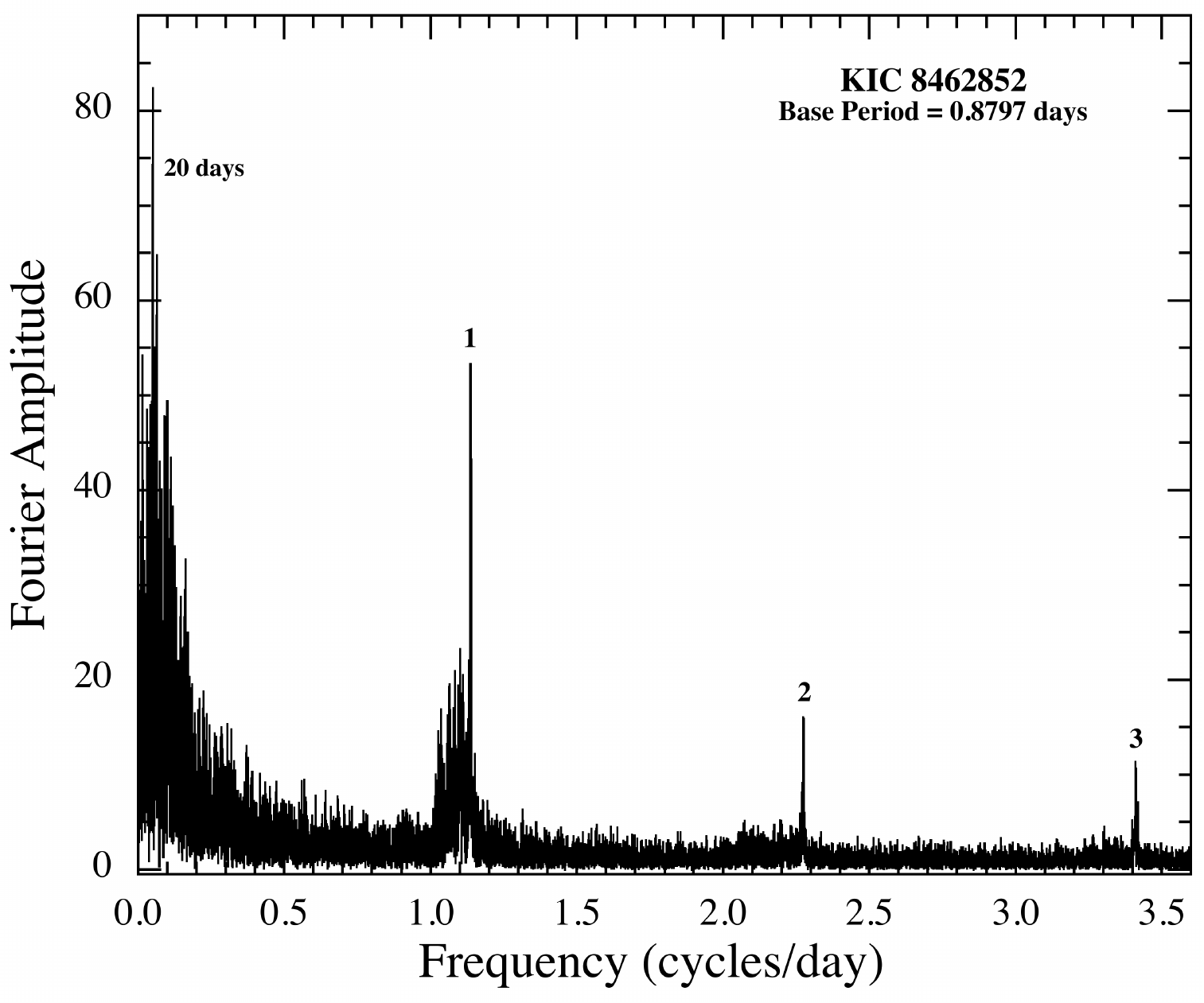}
    \caption{Fourier transform for \thisstar. The peaks are labeled with the harmonic numbers starting with 1 for the base frequency. Refer to Section~\ref{sec:kepler} for details.}
    \label{fig:fft}
  \end{figure}

There are also modulations in the raw flux data at the $\sim 500$~ppm level which are visible by eye.  To further explore whether any of these modulations are periodic, or have a periodic component, we generated a Fourier transform (FT) of the data with the dips excised from the data train.  Figure~\ref{fig:fft} shows the FT of the \kepler\ photometry and one can see a clear periodicity of 0.88~day (1.14~cycles/day) and its next two higher harmonics.

This 0.88-day signal is a broad feature that resembles typical FTs of Kepler targets for early type stars (\citealt{bal13}, see their figure~6).  If this is a rotation period, then the projected rotational velocity (from Section~\ref{sec:spectroscopy}) of $84 \pm 4$~\kms\ represents a minimum stellar radius of $\sim 1.46$~\rsun, consistent with the radius of an F-type star (also see Section~\ref{sec:spectroscopy}).  
Also seen in Figure~\ref{fig:fft} just to the left of the base frequency is a broad collection of smaller peaks.  This suggests that something more complicated than a single rotating surface inhomogeneity is producing the observed signal.

%%%%%%%%%%%%%%%%%%%%%%%%%%%%%%%%%%%%%%%%%%%%%%%%%%%%%%%%%%
% FFT plot 
%%%%%%%%%%%%%%%%%%%%%%%%%%%%%%%%%%%%%%%%%%%%%%%%%%%%%%%%%%

We investigate the stability of the frequencies observed in the FT by performing a Short-Term Fourier Transform (STFT), again clipping the data in the dipping regions.  In the STFT method, the data are broken up into ``short'' segments of 43~d.   This segment duration has been selected to optimize both time and frequency resolution.  The FT is computed and displayed vertically on the plot, and this is repeated as a function of time, with overlap in time segments to gain back some temporal resolution.

\begin{figure}
  \centering  
      \includegraphics[width=84mm]
      {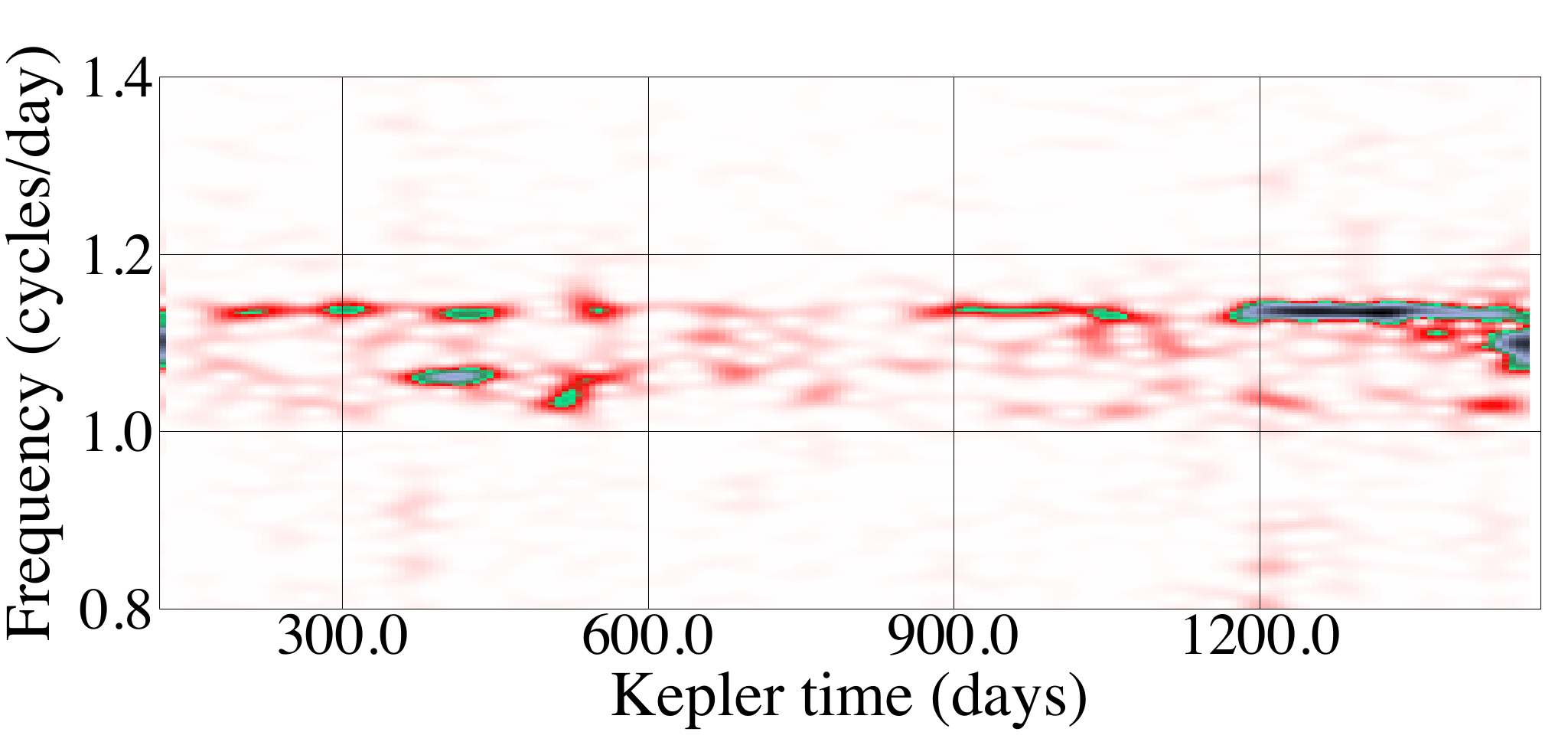}	
 \caption{The STFT for the \kepler\ flux time series.  The main base period of $\sim 0.88$~days is present throughout the span of observations. We identify (at least) two additional frequencies appearing around day 400 and 1400, corresponding to periods of 0.96 to 0.90~days, which we attribute to differential rotation.   Refer to Section~\ref{sec:kepler} for details. }
    \label{fig:stft}
  \end{figure}

The STFT is presented in Figure~\ref{fig:stft}.  This shows that the 0.88~day signal is present in most of the \kepler\ time series, with the strongest presence occurring around day 1200.  Interestingly however, around day 400 and day 1400, we see major contributions at different frequencies, corresponding to $\sim 0.96$~days and $\sim 0.90$~days, respectively.  We conclude that these are the source of the broad collection of peaks to the left of the base frequency noted above.  These low-frequency side-bands could possibly be due to regions contrasted in flux (e.g., starspots, chemically peculiar regions) appearing at higher latitudes coupled with differential rotation.  This is consistent with the differential rotation (or inferred fractional frequency difference of $\sim 10$\%) for F-type stars \citep{reinhold13}.  We would like to note however, that we cannot completely discount the possibility that these periods are due to pulsations. The position of \thisstar\ is within the Gamma Doradus ($\gamma$~Dor) region of the instability strip, where pulsations are observed at $< 5$~cycles~d$^{-1}$ \citep[e.g.,][]{uyt11}.  To investigate this, we then compared the STFTs of known $\gamma$~Dor pulsators to the STFT of \thisstar. We found that the dominant frequencies in STFTs for known $\gamma$~Dor stars do not evolve with time, contrary to the STFT for \thisstar.  This supports the interpretation that the $\sim 0.88$~d signal is due to the star's rotational period.  

%%%%%%%%%%%%%%%%%%%%%%%%%%%%%%%%%%%%%%%%%%%%%%%%%%%%%%%%%%
% STFT plot 
  
%%%%%%%%%%%%%%%%%%%%%%%%%%%%%%%%%%%%%%%%%%%%%%%%%%%%%%%%%%

\begin{figure}
  \centering  
 %\vspace{-0.3in}
       \includegraphics[trim=4mm 0mm 7mm 0mm, clip, width=84mm]%[width=84mm]
      {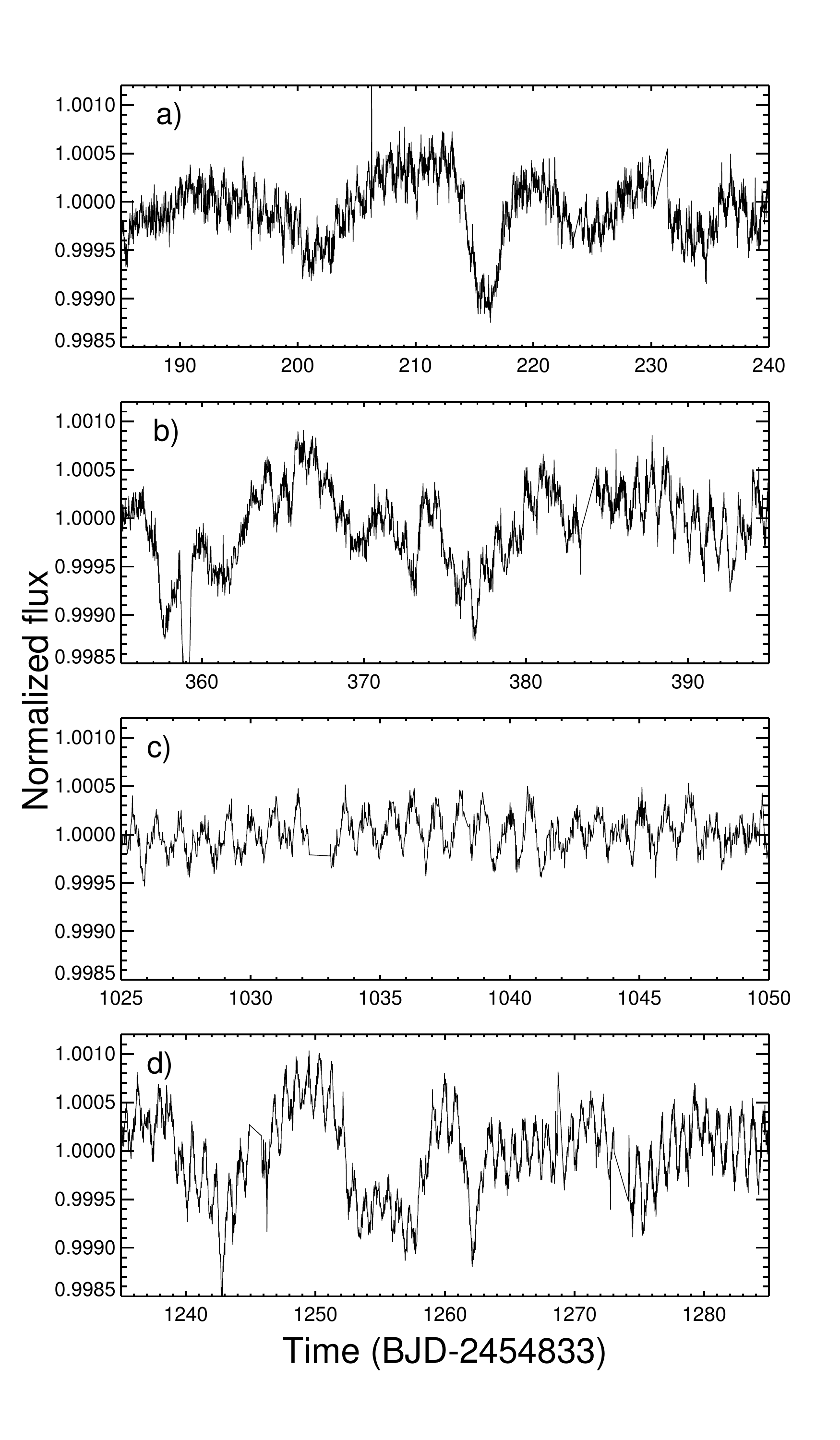}
 %\vspace{-0.4in}
    \caption{Stacked plots showing a zoomed-in portion of the \kepler\ light curve.  The star's rotation period of 0.88~d is seen in each panel as the high-frequency modulation in flux. With the exception of panel `c)', a longer term (10 --20~day) brightness variation is observed, also present in the FT shown in Figure~\ref{fig:fft}. Refer to Section~\ref{sec:kepler} for details.}
    \label{fig:kepler1}
  \end{figure}

We also report on the presence of variability on the timescale of 10 -- 20 days (Figure~\ref{fig:fft}), which, when present, is visible by eye in the light curve\footnote{Also present in the raw SAP data.}.  We illustrate this in Figure~\ref{fig:kepler1}, showing zoomed in regions of the \kepler\ light curve.  The star's 0.88~d period is also evident in each panel as the higher-frequency flux variations.  The panel second from the bottom `(c)' shows no low-frequency (10 -- 20 day) variations, but the rest do.  While the largest of the dipping structures within the D1500 events could also be described as having a periodic structure close to 20 days, the magnitude of the variability and the temporal behavior are much different than these low-amplitude variations described here. Thus, we cannot suggest any connection between the D1500 features and the 10-20 day variability.  Finally, we note that the 10 -- 20 day variability may actually arise on a faint neighboring star (see Sect.~\ref{sec:imaging}).  

%%%%%%%%%%%%%%%%%%%%%%%%%%%%%%%%%%%%%%%%%%%%%%%%%%%%%%%%%%
% closeup fluxes plot 
%%%%%%%%%%%%%%%%%%%%%%%%%%%%%%%%%%%%%%%%%%%%%%%%%%%%%%%%%%

%%%%%%%%%%%%%%%%%%%
%	Table of Dip Times
%%%%%%%%%%%%%%%%%%%

\begin{table}
 %\centering
 \caption{Principal Dip Times of KIC 8462852 vs. 48.4-day Period}
 \label{tab:Dip-Table}
\begin{tabular}{llcccc}
 \hline
 \hline
 \noalign{\smallskip}
dip \# & name & depth & BJD &  cycles & $|$residual$|$  \\
 & & & ($-$ 2\,454\,833)  & (from dip 5) &  (from integer)         \\
 \noalign{\smallskip}
 \hline
 \noalign{\smallskip}
 1 & (D140) &  0.5\% & 140.49    &   -13   &   0.52   \\  
 2 & (D260) &   0.5\% & 261.00   &   -11   &   0.01   \\  
 3 & (D360) &  0.2\% & 359.11    &   -9     &    0.04  \\ 
 4 & (D425) & 0.2\% &  426.62   &    -7    &    0.44  \\  
 5 & (D800) & 16\% & 792.74      &     0    &     0.00 \\
 6 & (D1200) & 0.4\% & 1205.96   &    8     &    0.54  \\
 7 & (D1500)  & 0.3\% &  1495.97 &  14     &    0.53  \\
 8  & (D1520) &  21\% & 1519.60  &  15     &    0.02  \\
 9  & (D1540)  & 3\% &1540.40      &  15     &   0.45  \\
 10 & (D1570) & 8\% & 1568.49    &   16     &   0.03  \\
 \noalign{\smallskip}
 \hline
\end{tabular}
 %\vspace{-0.1in}
\end{table}
%%%%%%%%%%%%%%%%%%%

There is another possible periodicity that is worth discussing briefly. In Table~\ref{tab:Dip-Table}, we summarize the times and depths of 10 discrete dips present in the \kepler\ light curve, also labeled in panel `(b)' -- `(e)' of Figure~\ref{fig:kepler}.  If we examine the two most prominent dips (D1568 and D1520; also see panel `(d)' in Figure~\ref{fig:kepler}), we see that they have a separation of $\sim 48.8$~days.  We can also see that the D800 dip (dip \#5 in Table~\ref{tab:Dip-Table}) is separated from the D1520 dip by 15 of these intervals, if the interval is more precisely defined to be 48.4 days.  Furthermore, the very shallow dips early in the \kepler\ time series at D260 and D360 are very close to 26 and 24 of these 48.4-day cycles from the D1520 dip.  The other five identified discrete dips (four of which are very shallow), also listed in Table~\ref{tab:Dip-Table}, are about a half cycle out of phase with this period to within $\sim \pm 5$\% of a cycle.  In this exercise, we have neglected the fact that the three most prominent dips in the D1500 region are quite highly structured, and they also have additional minima whose times could have been tagged and included in the analysis.  At this time we do not ascribe any particular significance to this period, but it is something to bear in mind as more data on this object become available.

%%%%%%%%%

\subsection{Spectroscopy}\label{sec:spectroscopy}

We obtained four high resolution ($R=47000$) spectra of \thisstar\ with the FIES spectrograph \citep{Frandsen1999,Telting2014} mounted at the 2.56-m Nordic Optical Telescope (NOT) of Roque de los Muchachos Observatory in La Palma, Spain. The observations were performed on 11 August 2014, 5 November 2014, 20 November 2015, and 26 November 2015. The data were reduced using standard procedures, which include bias subtraction, flat fielding, order tracing and extraction, and wavelength calibration. The extracted spectra have a S/N ratio of 45--55 per pixel at 5500~\AA. 

\begin{figure}
    \begin{center}
        \begin{tabular}{c}
            \includegraphics[trim=7mm 0mm 0mm 0mm, clip, width=84mm]
          {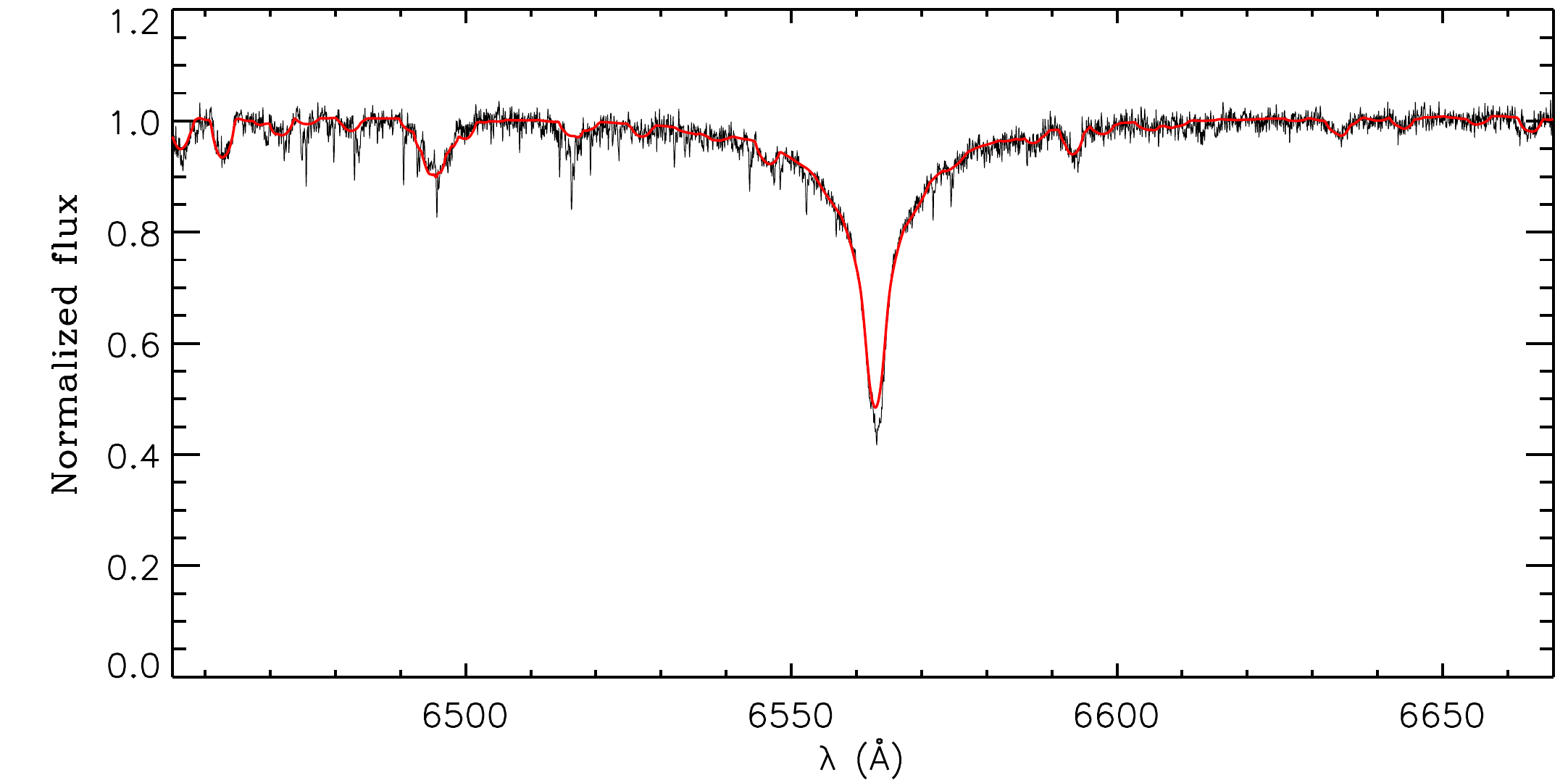}	\\
            \includegraphics[trim=7mm 0mm 0mm 0mm, clip, width=84mm]
            {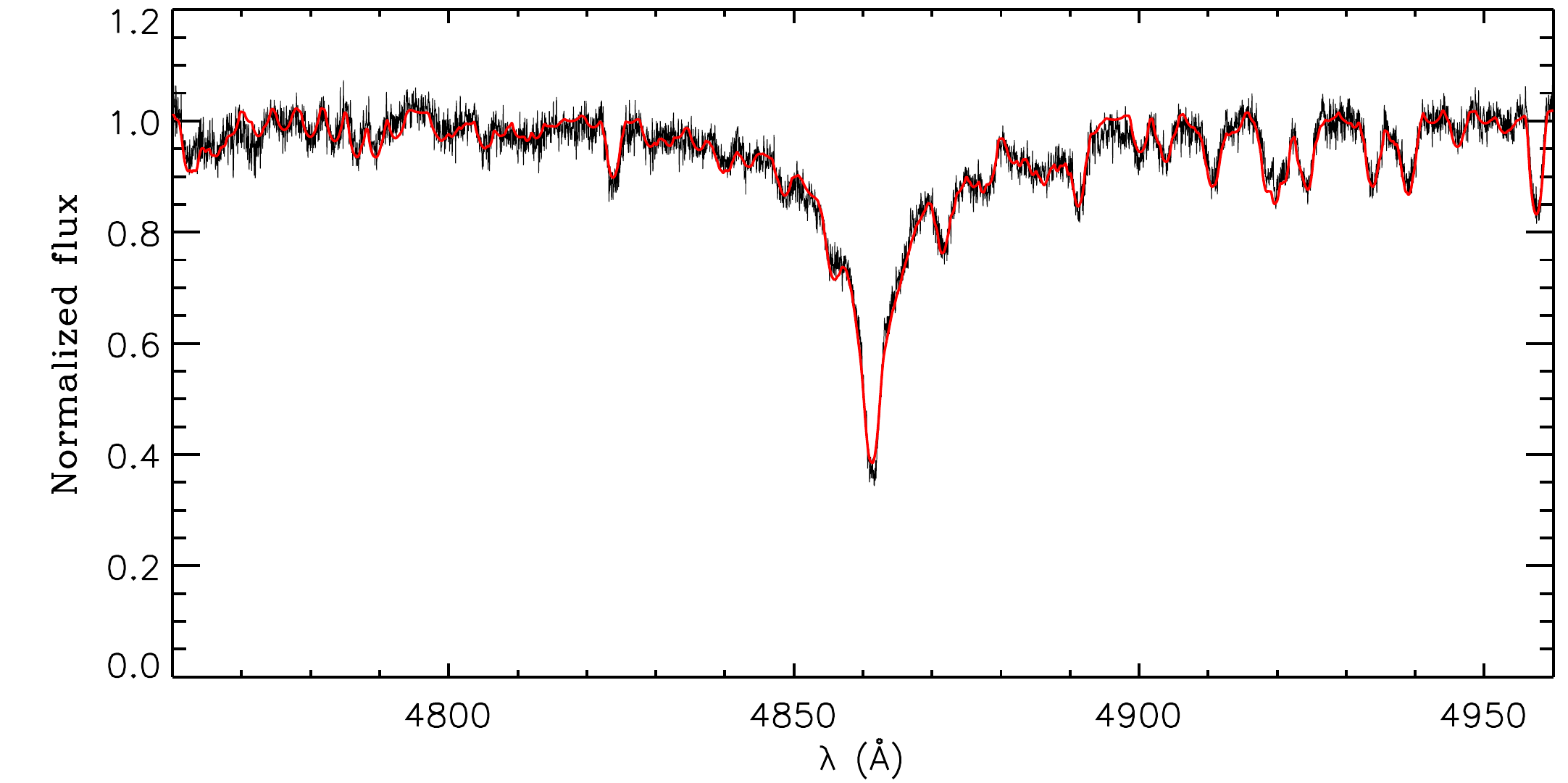}	\\
            \includegraphics[trim=7mm 0mm 0mm 0mm, clip, width=84mm]
            {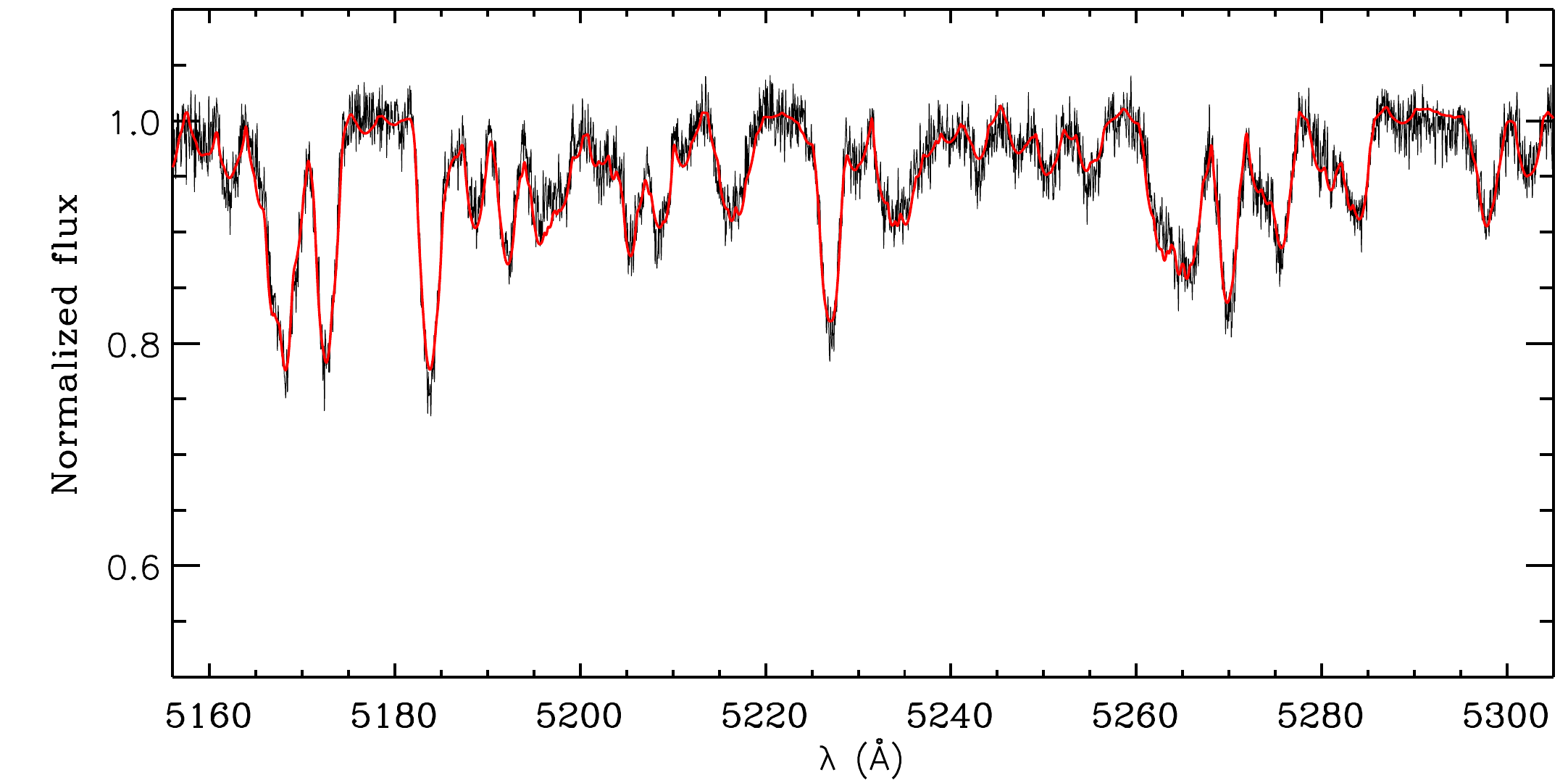}  \\
             \includegraphics[trim=7mm 0mm 0mm 0mm, clip, width=84mm]
             {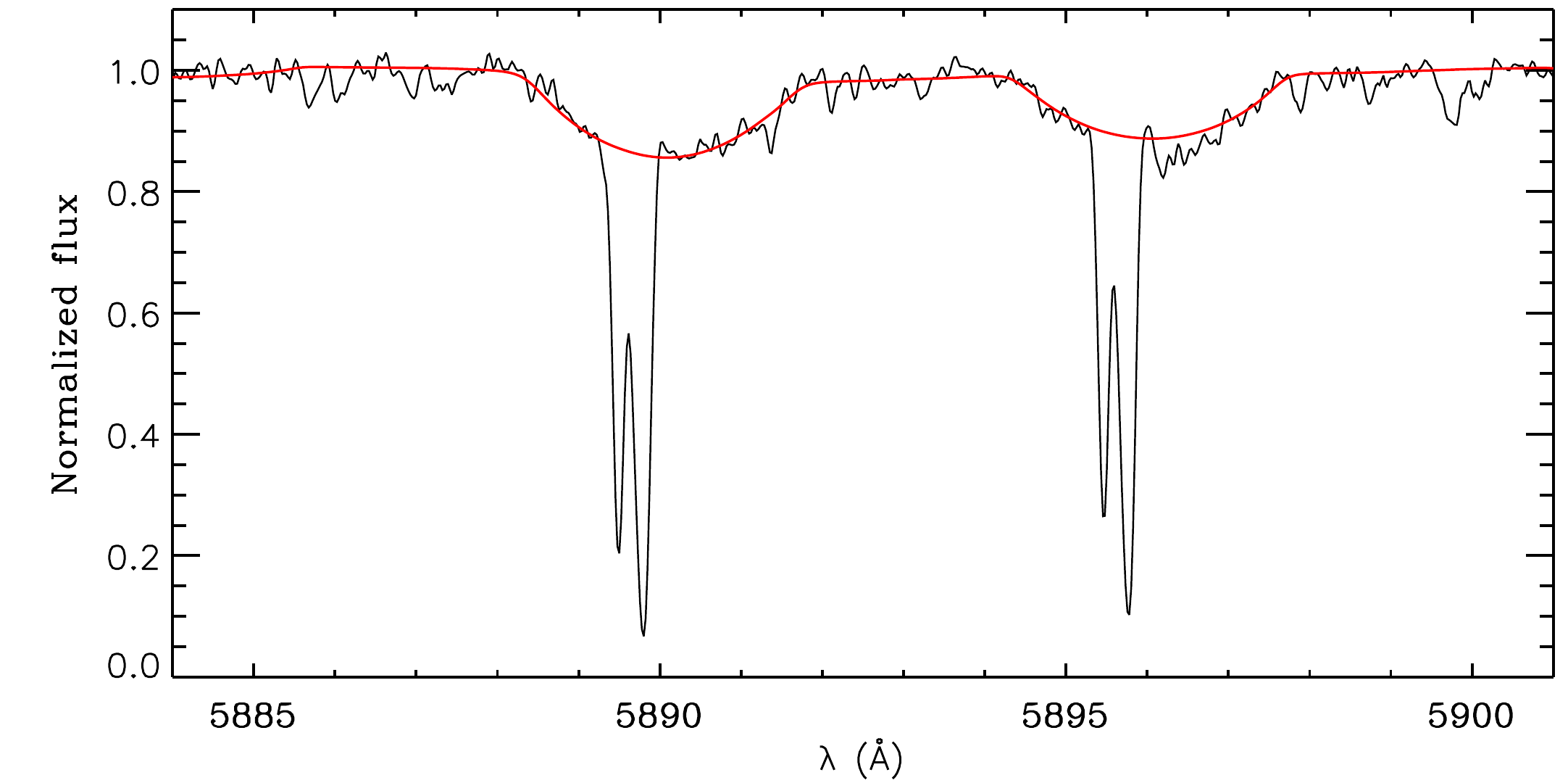}		           
        \end{tabular}
        \end{center}
    \caption{NOT spectrum closeups for \thisstar, the best fit stellar model shown in red.  Panels show region near H\,$\alpha$, H\,$\beta$, Mg, and Na D (top to bottom).  The bottom panel shows both the stellar (broad) and interstellar (narrow) counterparts of the Na D lines.  Refer to Section~\ref{sec:spectroscopy} for details.  }
    \label{fig:spec_montage}
  \end{figure}

Following the same spectral analysis procedure described in \citet{Rappaport2015}, we use the SPECTRUM code to calculate a grid of synthetic spectra using ATLAS9 models. We then use the co-added FIES spectrum to determine the stellar effective temperature $T_{\rm eff}$, surface gravity $\log g$, projected rotational velocity $v \sin i$, metal abundance [M/H], and spectral type of \thisstar\ (Table~\ref{tab:properties}). 
The plots in Figure~\ref{fig:spec_montage} show selected regions of the observed spectrum (black) along with the best fit model (red). 
The temperature we derive ($T_{\rm eff} = 6750\pm140$~K) is consistent with the photometric estimate of $T_{\rm eff} = 6584^{+178}_{-279}$~K from the revised Kepler Input Catalog properties \citep{hub14}, as well as with $T_{\rm eff} = 6780$~K derived from the empirical $(V-K)$ color-temperature relation from \citet{boy13a}. The projected rotational velocity we measure $v \sin i = 84 \pm 4$~\kms\ is also well in line with the one predicted from rotation in Section~\ref{sec:kepler}, if the 0.88~d signal is in fact the rotation period.  Overall, the star's spectrum is unremarkable, as it looks like an ordinary early F-star with no signs of any emission lines or P-Cygni profiles. Finally, we use the stellar properties derived from our spectroscopic analysis to estimate a stellar mass $M = 1.43$~\msun, luminosity $L = 4.68$~\lsun, and radius $R = 1.58$~\rsun, corresponding to a main-sequence F3~V star based on the empirical calibrations from \citet{pec13}\footnote{\url{http://www.pas.rochester.edu/\~emamajek/EEM\_dwarf\_UBVIJHK\_colors\_Teff.txt}}.  Combining the radius (assuming a conservative value of 20\% for the radius error), projected rotational velocity, and rotation period (Section~\ref{sec:kepler}), we determine a stellar rotation axis inclination of $68 \pm 29$~degrees.  

While interstellar medium features are not typically related to indicators of astrophysically interesting happenings in stars, we note the presence of stellar and interstellar Na~D lines in our spectra. In the bottom panel of Figure~\ref{fig:spec_montage}, we show a close up of the region containing the Na~D lines ($\lambda\lambda 5890,5896$\AA).  Within the two broad stellar features, there are two very deep and narrow Na~D lines with split line profiles, indicating the presence of two discrete ISM clouds with different velocities of $\sim 20$~\kms. 

%%%%%%%%%%%%%%%%%%%%%%%%%%%%%%%%%%%%%%%%%%%%%%%%%%%%%%%%%%
% spectroscopy montage plot 
%%%%%%%%%%%%%%%%%%%%%%%%%%%%%%%%%%%%%%%%%%%%%%%%%%%%%%%%%%

%%%%%%%%%
\subsection{Imaging}\label{sec:imaging}

\begin{figure}
  \centering  
      \includegraphics[width=84mm]
      {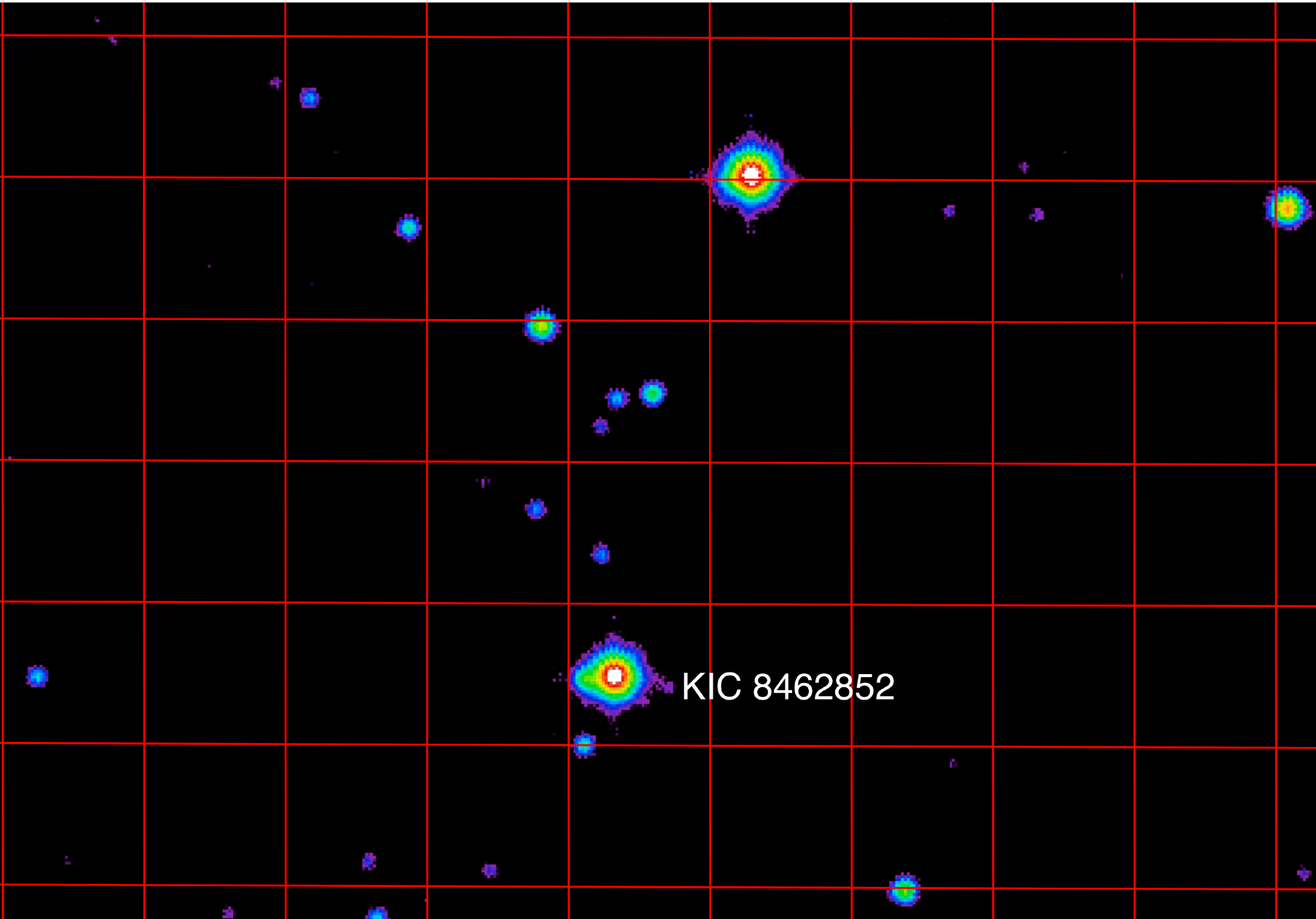} 
    \caption{UKIRT image for \thisstar\ and another bright star for comparison, showing that it has a distinct protrusion to the left (east). For reference, the grid lines in the image are $10^{\prime\prime} \times 10^{\prime\prime}$.  The color coding is logarithmically scaled. Refer to Section~\ref{sec:imaging} for details.}
    \label{fig:ukrit}
  \end{figure}

Figure~\ref{fig:ukrit} shows the UKIRT image of \thisstar\ as well as a similarly bright source $\sim 40^{\prime\prime}$ away.  The PSF of \thisstar\ is asymmetric by comparison, leading us to speculate that \thisstar\ has a faint companion star about $1.5-2^{\prime\prime}$ away.  

We observed \thisstar\ on UT 16 Oct 2014 using the natural guide star adaptive optics (AO) system \citep{wiz00} of the 10-meter Keck II Telescope on Mauna Kea, Hawaii. We used the facility IR camera NIRC2 and the $J$ (1.25~\micron), $H$ (1.64~\micron), and $K$ (2.20~\micron) filters from the Mauna Kea Observatories (MKO) filter consortium \citep{sim02, tok02}. We used NIRC2's narrow camera, which produces a 0.00994\arcsec~pixel\perone\ scale and a 10.2\arcsec\ field of view. Conditions were cloudy with variable seeing, around 1\arcsec\ FWHM. \thisstar\ was observed over an airmass range of 1.26--1.28.

\begin{figure}
  \centering  
      \includegraphics[width=84mm]
      {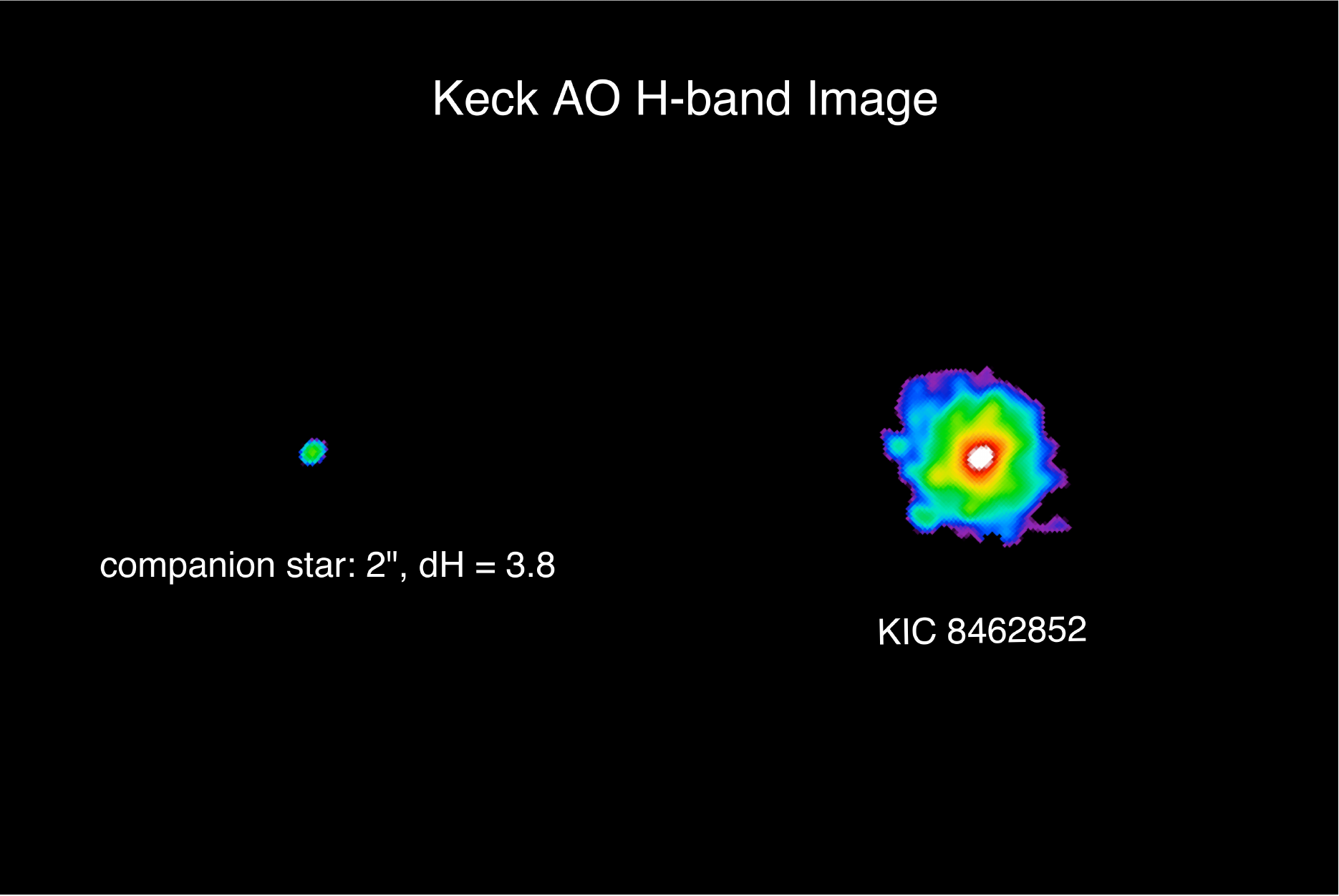} % stacked H band image
   \caption{Keck AO $H$-band image for \thisstar\ showing the companion was detected with a 2$^{\prime\prime}$ separation and a magnitude difference $\Delta H = 3.8$.  The color coding is logarithmically scaled. Refer to Section~\ref{sec:imaging} for details.}
    \label{fig:keckAO}
  \end{figure}

The AO-corrected images have full widths at half maxima (FWHMs) of 39~mas, 43~mas, and 51~mas at $JHK$, respectively, with RMS variations of about 1--3\%. We obtained a series of nine images in each filter. The total on-source integration time was 65~seconds per filter. The images were reduced in a standard fashion using custom scripts written in the Interactive Data Language (IDL). We constructed flat fields from the differences of images of the telescope dome interior with and without lamp illumination. We subtracted an average bias from the images and divided by the flat-field. Then we created a master sky frame from the median average of the bias-subtracted, flat-fielded images and subtracted it from the individual reduced images. The individual reduced images were registered and stacked to form a final mosaic (Figure~\ref{fig:keckAO}).

As suspected from the asymmetric UKIRT image, the Keck AO images reveal an obvious faint companion at a separation of 1.95\arcsec\ and position angle of 96.6\degs. To measure the flux ratios and relative positions of the two components, we used an analytic model of the point spread function (PSF) as the sum of two elliptical Gaussian components, a narrow component for the PSF core and a broad component for the PSF halo, as we have done for other binaries \citep{liu08}.  For the individual images obtained with each filter, we fitted for the flux ratio, separation, and position angle of the binary.  To correct for optical distortions in NIRC2, we used the calibration of \citet{yel10}.  The averages of the results were adopted as the final measurements and the standard deviations as the errors (Table~\ref{tab:properties}). 

It is unclear whether this is a physical or visual binary, though given the delta magnitude and separation, the chance alignment of the companion being a background or foreground object is only $\sim 1$\% \citep{rap14b}. At $\sim 2$\% of the flux of the brighter star, this would be a $\sim 0.4$~\msun~M2\,V star, if it is indeed at the same distance as our target F star \citep{kra07}.  The $JHK$ colors are also consistent with the companion being a dwarf, not a giant \citep{bes88}.  If we take the magnitude of \thisstar\ as $V = 11.705$, and the absolute visual magnitude of an F3V star to be $V = 3.08$ \citep{pec13}, then the (reddened) distance modulus is 8.625.  We derive a de-reddened distance of $\sim 454$\,pc using $E(B-V) = 0.11$ (Section~\ref{sec:sed}; corresponding to a $V$-band extinction of $A_V = 0.341$).  Assuming the fainter star is associated with the main F-star target, the angular separation of $\sim 1.95^{\prime\prime}$ translates to a distance of $\sim 885$~AU.  
At this separation, the second star cannot currently be {\it physically} affecting the behavior of the Kepler target star, though could be affecting bodies in orbit around it via long term perturbations \citep[see][]{kai13}.  If such a star is unbound from \thisstar, but traveling through the system perpendicular to our line of sight, it would take only 400 years to double its separation if traveling at $10$~km~sec$^{-1}$. So, the passage would be relatively short-lived in astronomical terms.

We also obtained Speckle observations of \thisstar\ on the night of UT 22 Oct 2015 using the DSSI instrument on the WIYN telescope located on Kitt Peak \citep{2011AJ....142...19H}. Observations were made simultaneously in two filters with central wavelengths at 692 and 880~nm. Both filters show the source to be single, with no visible companion observed to within 0.08~arcsec and brighter than a delta magnitude of 3.8 and 4.2 magnitudes (for the 692 and 880~nm filters, respectively). The companion star seen in the Keck NIRC2 image was not detected, favoring the conclusion that it is an M-dwarf, which would be too faint to be detected in the reddest DSSI filter (880~nm).  However, it is important to note that these speckle results provide an independent confirmation of the results from Keck AO: \thisstar\ has no additional companion down to a separation of $\sim 20$~AU detectable within the relative brightness limits with each instrument.

Finally, we speculate that the 10-20 day periodicity discussed in Sect.~\ref{sec:kepler} might actually arise on the neighboring faint M star.  The amplitude of those modulations are $\sim$500 ppm of the total target flux.  If they arise on the M-star, then their fractional modulation of that star would be as high as 3\%, which would not be unusual for an M star.

%%%%%%%%%%%%%%%%%%%%%%%%%%%%%%%%%%%%%%%%%%%%%%%%%%%%%%%%%%
% UKRIT plot 
%%%%%%%%%%%%%%%%%%%%%%%%%%%%%%%%%%%%%%%%%%%%%%%%%%%%%%%%%%

%%%%%%%%%%%%%%%%%%%%%%%%%%%%%%%%%%%%%%%%%%%%%%%%%%%%%%%%%%
% Keck AO plot 
%%%%%%%%%%%%%%%%%%%%%%%%%%%%%%%%%%%%%%%%%%%%%%%%%%%%%%%%%%

%%%%%%%%%
\subsection{Spectral energy distribution}\label{sec:sed}

\begin{figure}
  \centering  
  \includegraphics
         [trim=9mm 0mm 0mm 0mm, clip,width=85mm]
	{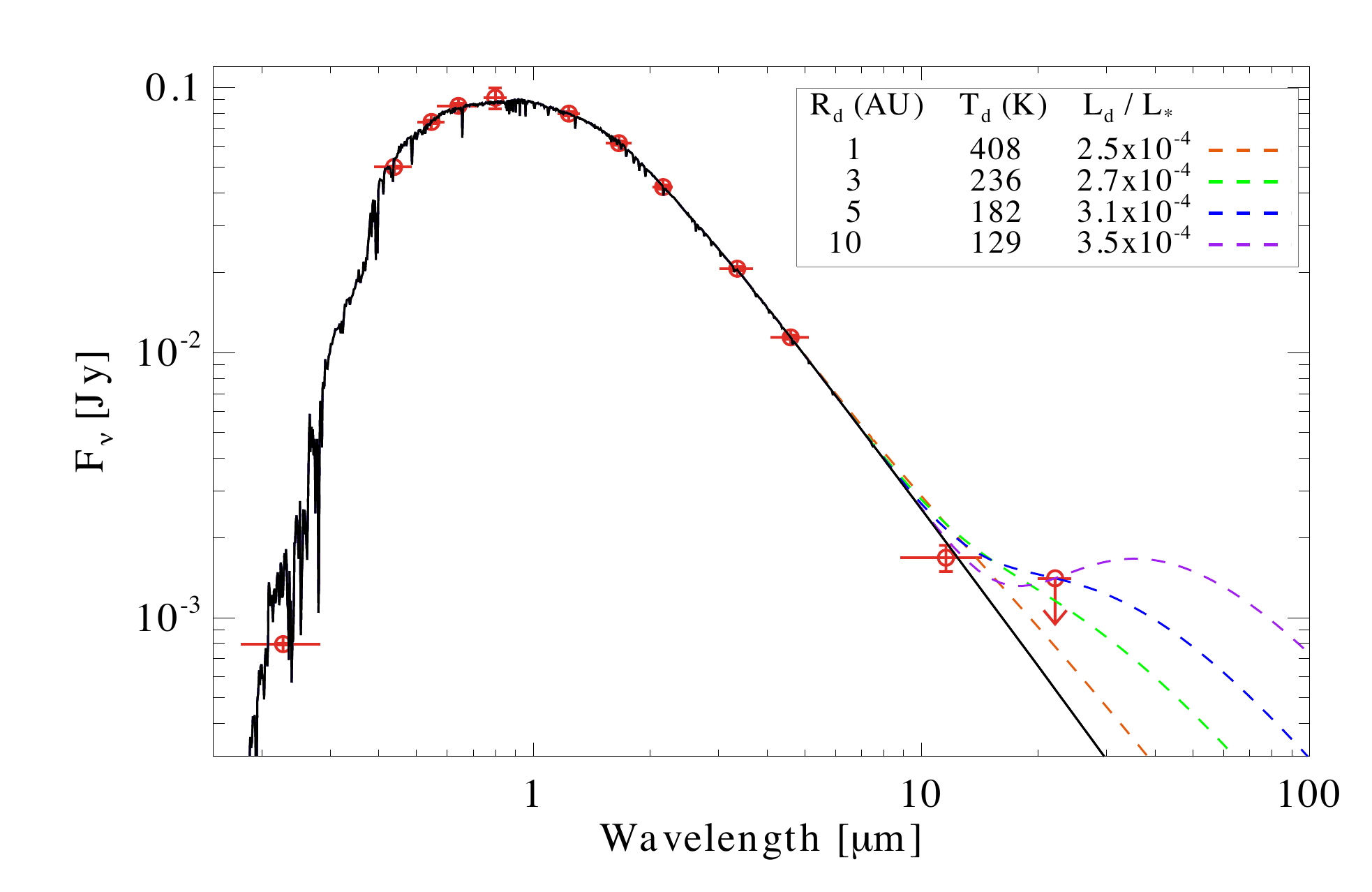}	
    \caption{SED for \thisstar. The Black solid line is a model for a star with $T_{\rm eff} = 6750$~K and $E(B-V) = 0.11$. Flux calibrated photometry are plotted in red, where the extent of the ``error-bars'' in the X-direction indicate the wavelength range of each bandpass and the Y-direction shows the error of the flux measurement.    
     Refer to Section~\ref{sec:sed} for details.} 
    \label{fig:sed}
  \end{figure}

The spectral energy distribution (SED) of KIC 8462852 including optical, 2MASS \citep{skr06}, (ALL)WISE \citep{wri10}, and Galex NUV \citep{Morrissey07} flux densities is shown in Figure~\ref{fig:sed}.  Optical photometry in $BV(RI)_C$ filters was obtained by the 90\,cm Schmidt telescope of the Konkoly Observatory at Piszk\'estet\H{o} Mountain Station. For standard magnitudes GD391\,ABCE photometric standard stars were used as comparison \citep{lan13}. Photometric magnitudes are listed in Table~\ref{tab:properties}.  

In order to study whether the system exhibits excess at mid-infrared wavelengths, we first fitted an {\sc ATLAS9} atmosphere model \citep{cas04} to the photometric data points between 0.15 and 3.6{\micron}. From the grid of model atmospheres we selected the one that has the closest metallicity, surface gravity, and effective temperature to those derived from our spectroscopic study. Thus we fixed $T_{\rm eff}$, $\log{g}$, and [Fe/H] parameters to 6750~K, 4.0, and 0.0, respectively, and only the amplitude of the model and the reddening were fitted. The best fitted photospheric model is displayed in Figure\,8.  We derive a  reddening of $0.11 \pm 0.03$~mags.  By comparing the measured $W2$ and $W3$ WISE flux densities at 4.6 and 11.6~{\micron} (at 22~{\micron} we have only an upper limit) with the predicted fluxes derived from the photosphere model we found them to be consistent, i.e., no excess emission can be detected at mid-infrared wavelengths.  This lack of significant IR excess is independently confirmed using warm Spitzer/IRAC data by \citet{2015ApJ...814L..15M}.

%%%%%%%%%%%%%%%%%%%%%%%%%%%%%%%%%%%%%%%%%%%%%%%%%%%%%%%%%%
% SED plot 
%%%%%%%%%%%%%%%%%%%%%%%%%%%%%%%%%%%%%%%%%%%%%%%%%%%%%%%%%%

However, this does not exclude the existence of a colder debris disk or a warmer, but relatively tenuous disk. Assuming that the emitting grains act like a blackbody, we can derive their characteristic temperature at a specific stellar-centric distance. Using this approach, we compute the SED of a narrow dust belt located at a distance of 1, 2, 3, 5, and 10~AU from a star with a luminosity of 4.7~\lsun, corresponding to the main-sequence stage \citep{pec13}. The $W3$ and $W4$ band photometry 
were then used as upper limits to set the amplitude of the excess.  
Figure~\ref{fig:sed} shows the result of these computations and summarizes the fundamental disk properties (dust temperature, upper limits for fractional luminosity) of the dust belts at different radii.  It is worth noting that this very simple model accounts only for large blackbody grains, smaller ($\mu$m-sized) grains are ineffective emitters and may be heated to higher temperatures compared to larger grains at the same location.  We revisit this analysis in more detail later in Section~\ref{ss:indep} (also see Figure~\ref{fig:detlim}).

%%%%%%%%%

\subsection{Ground-based photometric surveys}

We reviewed the $\sim 700$ photometric intensities from the years 1900 -- 2000 from the Digital Access to a Sky Century \@ Harvard (DASCH) project\footnote{\url{http://dasch.rc.fas.harvard.edu/index.php}} \citep{gri12}.  The error bars on the photometry are about $\sim 10$\%.  At this level, we found the star did not do anything spectacular over the past 100 years.  However, if it underwent several $\sim 20$\% dips in flux lasting for several days each during that period, the chances are high that there were no plates exposed at those times.  

{\it SuperWASP} data \citep{but10} are unremarkable for \thisstar.  We note that there is a 0.2 magnitude offset between the available {\it SuperWASP} data sets. 
However, we see the same offset when comparing its photometry with a similarly bright source nearby \thisstar.  Thus, we reject this being real (e.g., due to a flaring event, etc.).  

Unfortunately, \thisstar\ falls outside the area covered by the {\it KELT} network (T. Beatty, private communication).

%%%%%%%%%

\subsection{Limits on a close companion}\label{sec:elv}

We use the four FIES spectra (Section~\ref{sec:spectroscopy}) to measure the presence of any Doppler shifts induced by a companion.  We traced the radial velocity (RV) drift of the instrument by taking long-exposed ThAr spectra in a bracketed sequence, i.e., right before and after each target observation. RV measurements were derived by cross-correlating the target spectra with the rotationally broadened best fitting Kurucz model.  The RV measurements are listed in Table~\ref{tab:RV-Table} along with the error bars and the barycentric Julian dates in barycentric dynamical time. To within the $\sim 400$~m~s$^{-1}$ uncertainties in the RV measurements, the four values we measure are quite consistent with no change at all over the 470 day observation interval.  

\begin{figure}
  \centering  
      \includegraphics[width=84mm]
      {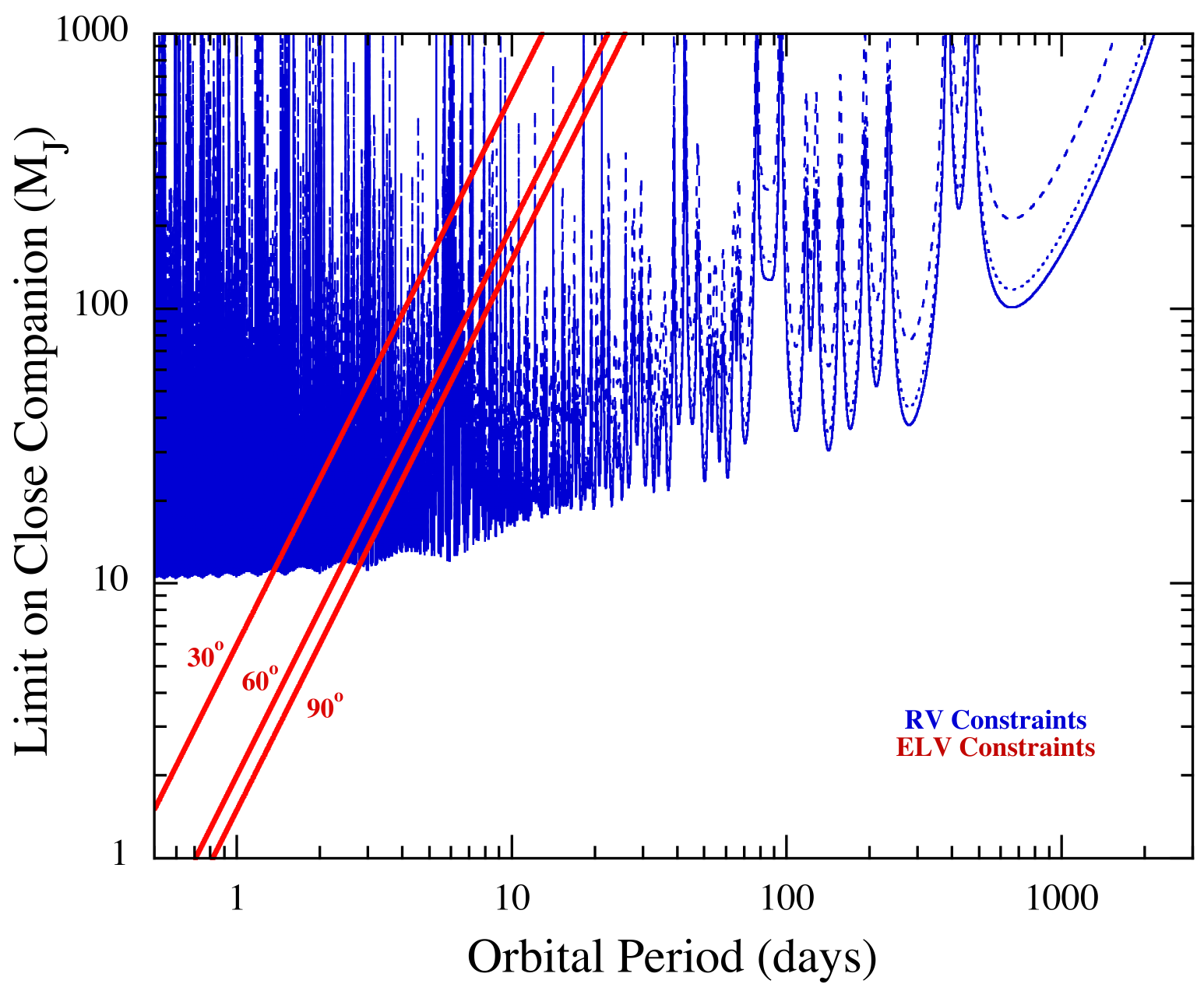}
 % \vspace{0.9in}
    \caption{Upper limits (2-$\sigma$ confidence) to a hypothetical companion mass from the lack of ELVs (red curves) and lack of RV variations on four occasions (blue curves).  Each type of constraint is shown for three different assumed orbital inclination angles ($30^\circ$, $60^\circ$, $90^\circ$); these are marked directly on the ELV constraint curves, and can be inferred from the dashed, dotted, and solid curves, respectively, for the RV constraints. These results indicate that there are no objects heavier than a super-Jupiter in close orbits with $P_{\rm orb} \lesssim 2$ days, and likely no heavier in mass than a brown dwarf for $P_{\rm orb} \lesssim 300$ days.  Refer to Section~\ref{sec:elv} for details.}
    \label{fig:Mass_Constraints} 
  \end{figure}

In order to quantify what limits we can set on the mass of an hypothetical close companion star, we carried out the following analysis.  We assumed a circular orbit because there are insufficient data points to fit for the parameters in an eccentric orbit.  Then, for each in a sequence of $4 \times 10^5$ trial orbital periods, $P$, in the range of 0.5 to 3000 days, we fit the four RV points with a sine and cosine term to represent the orbit and a systemic $\gamma$ velocity.  From this fit we computed the velocity semi-amplitude K and added its 2-$\sigma$ uncertainty to establish a conservative upper limit to K. We then used the upper limit on K to compute the corresponding upper limit on the mass function.  Finally, we solved for the upper limit on the mass of the hypothetical close companion by taking the mass of the F star to be 1.4 $M_\odot$, and assuming three different orbital inclination angles ($30^\circ$, $60^\circ$, and $90^\circ$). The results are shown in Figure~\ref{fig:Mass_Constraints}.  The spikes are at values of $P$ where the epochs of the four RV measurements are commensurate with being at orbital phase 0, and the mass constraints are weaker at these periods.  For longer periods, the density of these spikes diminishes greatly and the lower locus of points can be taken as a likely upper limit on the mass of any companion. Therefore, we conclude that for periods between $\sim$30 and 300 days, the mass of any companion is very unlikely to exceed that of a brown dwarf. 

%%%%%%%%%%%%%%%%%%%
%	Radial velocity table
%%%%%%%%%%%%%%%%%%%

\begin{table}[t]
 \centering
 \caption{FIES RVs of \thisstar}
 \label{tab:RV-Table}
\begin{tabular}{ccc}
 \hline
 \hline
 \noalign{\smallskip}
BJD &   RV    & $\sigma_{\mathrm RV}$  \\
($-$ 2\,450\,000)  &  [km/s] &    [km/s]              \\
 \noalign{\smallskip}
 \hline
 \noalign{\smallskip}
    6881.5183     &   4.160   &   0.405 \\  % 4697
    6966.3635    &    4.165   &   0.446 \\  % 0435
    7347.3856     &   3.817  &    0.406 \\ % 0823
    7353.3511    &    4.630  &    0.436 \\  %  5147
 \noalign{\smallskip}
 \hline
\end{tabular}
\\ 
See Section~\ref{sec:elv} for details.
\end{table}
%%%%%%%%%%%%%%%%%%%

Another diagnostic to constrain the nature of the companion uses the FT in Figure~\ref{fig:fft}, which shows no sharp, narrow peaks without harmonics (Section~\ref{sec:kepler}).  With this information, a very basic limit can be set on a companion from the lack of observed ellipsoidal light variations (ELVs). The ELV amplitude $A_{\rm ELV}$ is expressed as: 
\begin{equation}
	A_{\rm ELV} \simeq 1.5 (M_c/M_{\ast}) (R_{\ast}/a)^3 \sin^2 i
\end{equation}
\noindent (e.g., \citealt{kop59,car11b}) where $M_{\ast}$ and $R_{\ast}$ are the mass and radius of the primary, $a$ and $i$ are the semimajor axis and orbital inclination, and $M_c$ is the mass of a putative companion.  Rearranging to express $a$ as the orbital period $P$ using Kepler's third law, this equation simplifies to: 
\begin{equation}
	A_{\rm ELV} \simeq 3.3 \times 10^{-5} (M_c/{\rm M_J}) ({\rm 1 d}/P)^2 \sin^2 i 
\end{equation}
where now the companion mass $M_c$ is expressed in Jupiter masses $M_{\rm J}$ and $P$ is in days.  If ELVs were present, we would have seen a peak $\gtrsim 50$~ppm for all periods shorter than 4 days ($\gtrsim 0.25$~cycles\,day$^{-1}$) in the FT (Figure~\ref{fig:fft}).   

The limits on the companion mass that we can set from the lack of ELVs, as a function of orbital period, are illustrated in Figure~\ref{fig:Mass_Constraints}.  They are plotted as red lines for three different assumed inclination angles. Note that an angle of 90$^\circ$ is not allowed or we would have seen (regular) transits; it is shown in this figure for instructive purposes only. These ELV mass constraints are superposed on those discussed above based on the lack of RV differences among our four measurements (blue curves). Taken together, these results indicate that there is not likely to be a close companion to the F star more massive than a super-Jupiter with $P_{\rm orb} \lesssim 2$ days, nor more massive than a brown dwarf for $P_{\rm orb} \lesssim 300$ days.  

%%%%%%%%%%%%%%%%%%%%%%%%%%%%%%%%%%%%%%%%%%%%%%%%%%%%%%%%%%
% Figure: Mass constraints 
%%%%%%%%%%%%%%%%%%%%%%%%%%%%%%%%%%%%%%%%%%%%%%%%%%%%%%%%%%

%%%%%%%%%
\subsection{Space motion and age}\label{sec:motion}
%%%%%%%%%
Using our distance estimate of 454\,pc (Section~\ref{sec:imaging}), the radial velocity obtained from the FIES spectrum (Section~\ref{sec:elv}), and proper motions and positions from the UCAC4 catalogue we computed the Galactic space motion of the target, yielding $+31.5$, $-2.5$, and $+10.2$~\kms\ for the U (defined as positive toward the Galactic center), V, and W velocity components, respectively. Young disk population stars have low velocity dispersion and they occupy a special region within the velocity space. Based on the studies of \citet{egg89}, \citet{leg92} defined a box by $-50<$~U~$<+20$~\kms, $-30<$~V~$<0$~\kms, and $-25<$~W~$<10$~\kms, which includes most of the young disk stars in our neighborhood. Our target lies outside of this box. In fact, its galactic space motion -- especially the U component -- deviates significantly from the characteristic space motion of any nearby young ($<100$~Myr) kinematic groups, open clusters, and star forming regions \citep{2007ApJS..169..105M,2015arXiv150706697M}.  Altogether, it implies that \thisstar\ likely does not belong to the youngest stellar population. 

In making this distance estimate, we assumed that \thisstar\ is a main-sequence star (Section~\ref{sec:imaging}). We note that assuming a pre-main or post-main sequence phase does not change our previous conclusion. These evolutionary stages would be accompanied by larger luminosities and thereby larger distances.  This would result in a galactic space motion that deviates even more significantly from that of typical young disk stars.  Unfortunately, our star falls outside the region where empirically calibrated age diagnostics such as chromospheric activity or stellar rotation period can be used (e.g., \citealt{mam08}).      

%%%%%%%%%%%%%%%%%%%%%%%%%%%%%%%%%%%%%%%%%%%%%%%%%%%%%%%%%%%%%%%
%               TABLE x  : properties
%%%%%%%%%%%%%%%%%%%%%%%%%%%%%%%%%%%%%%%%%%%%%%%%%%%%%%%%%%%%%%%

\begin{table}
\tabletypesize{\scriptsize}
\centering
\caption{Properties of \thisstar \label{tab:properties}}
\begin{tabular}{@{}rcc@{}}
\hline
Property & 
Value    &
Method/Reference  \\
\hline

RA  (deg)           &   \phn\phn\phn 301.564392      &  KIC \\
DEC (deg)           &   \phn\phn\phn\phn 44.456875   &  KIC \\
$K_p$ (mag)         &    \phn11.912                 &   KIC \\
$B$ (mag)	&	$12.262 \pm 0.008$\phn	&	90\,cm Schmidt (\S~\ref{sec:sed})	\\
$V$ (mag)	&	$11.705 \pm 0.017$\phn	&	90\,cm Schmidt (\S~\ref{sec:sed})	\\
$R_{\rm C}$ (mag)	&	$11.356 \pm 0.024$\phn	&	90\,cm Schmidt (\S~\ref{sec:sed})	\\
$I_{\rm C}$ (mag)	&	$11.051 \pm 0.098$\phn	&	90\,cm Schmidt (\S~\ref{sec:sed})	\\
$J$ (mag)           &    $10.763 \pm 0.021$\phn     &   2MASS \\	
$H$ (mag)           &    $10.551 \pm 0.019$\phn	    &   2MASS \\
$K$ (mag)           &    $10.499 \pm 0.020$\phn     &   2MASS \\
$W1$ (mag)           &   $10.425 \pm 0.023$\phn    &   (ALL)WISE \\
$W2$ (mag)           &   $10.436 \pm 0.020$\phn     &  (ALL)WISE \\			
$W3$ (mag)           &   $10.591 \pm 0.123$\phn      &   (ALL)WISE \\
$W4$ (mag)           &  \phn\phn $9.423$\tablenotemark{a}            &   (ALL)WISE \\
Rotational period (d)   &  $0.8797 \pm 0.0001$                & FT (\S~\ref{sec:kepler})     \\
Spectral type	&  F3 V   &  Spectroscopy (\S~\ref{sec:spectroscopy})         \\	
$T_{\rm eff}$ (K)   &   $6750 \pm 120$\phn &   Spectroscopy (\S~\ref{sec:spectroscopy})    \\
$\log g$ (cgs)      &   $4.0 \pm 0.2$ &   Spectroscopy (\S~\ref{sec:spectroscopy})    \\
$[$M/H$]$ (dex)	&	$0.00 \pm 0.10$	& Spectroscopy (\S~\ref{sec:spectroscopy})    \\		
$v \sin i$ (\kms)      &   $84 \pm 4$\phn &   Spectroscopy (\S~\ref{sec:spectroscopy})    \\
distance (pc)		&	454	&	Distance modulus (\S~\ref{sec:imaging})	\\
$E(B-V)$ (mag)            &   $0.11 \pm 0.03$ &   SED (\S~\ref{sec:sed})    \\

\hline

Binary separation (arcsec)		&	\phn 1.96	 &	Keck AO (\S~\ref{sec:imaging})		\\
Binary position angle (deg)		&	 96.6 \phn	 &	Keck AO (\S~\ref{sec:imaging})		\\
$\Delta J$ (mag)	&	$4.209 \pm 0.044$		&	Keck AO (\S~\ref{sec:imaging})		\\
$\Delta H$ (mag)	&	$3.840 \pm 0.017$		&	Keck AO (\S~\ref{sec:imaging})		\\
$\Delta K$ (mag)	&	$ 3.619 \pm 0.012$		&	Keck AO (\S~\ref{sec:imaging})		\\
%Spectral type

\hline
\end{tabular}
\tablenotetext{a}{Upper limit.}
\end{table}
%%%%%%%%%%%%%%%%%%%%%%%%%%%%%%%%%%%%%%%%%%%%%%%%%%%%%%%%%%%%%%%%
\subsection{Similar dippers in the \kepler\ field?}\label{sec:kepler_dippers}
The anomalous dips in \thisstar\ were serendipitously found by the Planet Hunter citizen science group.  Due to its aperiodic nature, it likely never would have been flagged/recovered by most searches for transits, eclipsing binaries, or asteroseismologically interesting stars.  However, knowing the existence of \thisstar's light curve, we naturally wondered if there are, in fact, numerous other such objects in the main-field \kepler\ data base.  We therefore applied a simple algorithm to search the data base for other systems similar to \thisstar.  The algorithm consisted of searching for dips with depths of greater than 10\% (i.e., normalized fluxes of $< 0.9$) that consist of 5 or more consecutive \kepler\ long-cadence samples (i.e, lasting more than $\sim 2.5$~hours).  In all, this search turned up more than a thousand targets with this signature.  The vast majority of them, however, were due to (1) eclipsing binaries, (2) the rotation signature of large amplitude starspots, and (3) some obvious \kepler\ data artifacts.  We carefully examined the remaining small number of systems by eye, but could identify none that was reminiscent of \thisstar.  We also lowered the threshold for dips to 5\%, but the search likewise turned up no candidates that one would believe closely resemble \thisstar.  Of course, some of the visual comparison work is necessarily qualitative, but we were satisfied that there are at most a few similar systems to be found in the main \kepler\ field.

%%%%%%%%%%%%%%%%%%%%%%%%%%%%%%%%%%%%%%%%%%%%%%%%%%%%%%%%%%%%%%%%

\section{Possible explanations of the observed dipping events observed in \thisstar}\label{sec:features}

The main issue in explaining the peculiar light curve for \thisstar\ is related to the presence of multiple dimming events, that are not periodic and of which the D800 single event has a smooth, yet highly asymmetric, profile, and the D1500 events are the deepest and most complex.  Here, we introduce several scenarios to explain \thisstar\ and discuss how the observational data do and do not support each theory. 

\subsection{Instrumental effects or data reduction artifacts?}\label{sec:instrumental}

The \kepler\ light curve for \thisstar\ is unique, and we have thoroughly explored the raw data for defects/instrumental effects, which could cause the observed variations in \thisstar's flux.  We use the \textsc{PyKe} software tools for Kepler data analysis to check the data for instrumental effects. We check the following possibilities:

\begin{itemize}
\item We checked that the same flux variations, i.e., the `dips', are present in the SAP$\_$FLUX data set.  
\item  We verified that data gaps and cosmic rays events\footnote{The times of these events are recorded in the headers of the fits files} do not coincide with the dipping events, as they are prone to produce glitches in the corrected fluxes.     
\item We verified at the pixel-level that there are no signs of peculiar photometric masks used in making the light curves.
\item We verified at the pixel level that the image light centroid does not shift during the `dipping' events  
\item We inspected the \kepler\ light curves of neighboring sources and find that they do not show similar variability patterns in their light curves.   
\item We determined that CCD cross talk, reflection, and column anomaly cannot be the cause \citep{cou14}. 
\item We verified with the \kepler\ team mission scientists that the data were of good quality.  
\end{itemize}

This analysis concludes that instrumental effects or artifacts in the data reduction are not the cause of the observed dipping events, and thus the nature of \thisstar's light curve {\bf is astrophysical in origin.}

\subsection{Intrinsic variability?}\label{sec:intrinsic_variability}

An example of a class of stars which display intrinsic variability are the R Coronae Borealis (RCB) type variables.  These are highly evolved F--G supergiants (e.g., \citealt{clayton96}) that have light curves which show pulsations (on the order of months) and irregular deep dips (lasting weeks to months).  Their ``dipping'' variability is associated with formation of clouds that obscure the photosphere, and is often observed as a sharp decrease in flux followed by a more gradual, and sometimes staggered, recovery.  In the case of \thisstar\, the time scales of the dips are different than those of a RCB variable.  Likewise, the ingress at D800 has a gradual decrease in flux, which is inverse to what is expected in a RCB, and the dip shapes at D1500 are also non-characteristic of a RCB.  Lastly, the spectroscopic $\log g$ and $v \sin i$ are far from those of a supergiant.  These items together strongly rule out the possibility of \thisstar\ being a RCB variable.     

Another possibility is the self-emission of disk material from the star itself, as in the case of Be-stars. Be stars are rapidly rotating (almost near breakup) stars that are usually of spectral class O and B, but sometimes A, and exhibit irregular episodic outbursts.  Usually these outbursts are in emission, but in some cases it can also result in dimming (see \citealt{hub98}).  Be stars also often exhibit quasi-periodic oscillations in the range of $\sim 0.5-1.5$ days.  This also fits the bill for what we see in the FT of \thisstar\ (\S~\ref{sec:kepler}).  It has been hypothesized \citep[e.g.,][]{rap82} that most, if not all, Be stars have a binary companion which originally transferred mass to the current Be star to spin it up to near breakup (the remnant of that star is sometimes found to be a neutron star).  The periods of these binaries range from a couple of weeks to thousands of days (perhaps longer).  If \thisstar\ is a Be star, we would get an unprecedented look into the inner disk behavior. In such as case, the broad peak in the FT at frequencies just below the 0.88~d periodicity could be explained by ejected material in a so-called ``excretion disk'' that is moving outward but with roughly Keplerian velocity.  

The lack of observed IR excess does not support the existence of an excretion disk. There is also an absence of H$\alpha$ emission in the star's spectrum (although, as noted above, Be star H$\alpha$ emission is known to be variable and turn off and on with timescales from days to years).   Furthermore, the temperature of KIC 8462852, $T_{\rm eff}$ = 6750 K, is too cool to be a Be star. It is also unlikely to have been spun-up by a donor star whose remnant is still orbiting the F star because of the constraints set by the four RV measurements and the limits on any ELVs (see Section~\ref{sec:elv}). Though, we cannot rule out remnants orbiting with $P \gtrsim$ a few years.

\subsection{Extrinsic variability?}\label{sec:extrinsic_variability}

\subsubsection{Related to the secondary star}

We first consider whether \thisstar's flux is contaminated by the nearby M-dwarf detected with high-resolution images (\S~\ref{sec:imaging}).  Whether or not the system is bound, the faint companion contributes light in the \kepler\ photometric aperture, which in turn affects the observed signal in the light curve. Our observations show that the flux ratio in the infrared is $\sim 30$, which translates to a factor of several hundred in the \kepler\ bandpass.  Thus, the maximum imprint that the M-dwarf has on the light curve variability is $\sim$$30$~mmags; this is insufficient to make an impression on \thisstar's light curve at anything greater than $\sim$3\%, and, in particular, it couldn't possibly explain any of the large dips.

%%%%%%%%%%%%%%%%%%%%%%%%%%%%%%%%%%%%%%%%%%%%%%%%%%%%%%%%%%%%%%%%

\subsubsection{Occultation by circumstellar dust clumps}

The dips could be readily explained in terms of occultation by an inhomogeneous circumstellar dust distribution. 
However, this does not mean that the dust distribution that would be required to explain the observations is physically plausible
or would necessarily apply to \thisstar.

Inhomogeneous dust distributions have been invoked to explain dips seen towards
some young stars such as UX Orionis or AA Tau-like ``dipper'' systems
\citep[][]{1994AJ....108.1906H,1999AJ....118.1043H,2009ApJ...699.1067M,cod14, ans15}.  
At an age of only a few tens of Myrs, these dipper stars have  $V$-band light curves characterized by sporadic 
photometric minima with amplitudes of 2 -- 3 magnitudes and with durations of days to many weeks. These objects 
also generally exhibit strong infrared excess, starting at $\sim 2-5$~$\mu$m and show signs of accretion (emission) in their spectra. 
However, in contrast to such systems, \thisstar\ has no detectable IR excess or accretion signature to suggest that it is a
young T Tauri star (Sections~\ref{sec:spectroscopy}, \ref{sec:sed}).
Thus a scenario in which material in a gas-dominated protoplanetary disk occults the
star due either to accretion columns or non-axisymmetric azimuthal or vertical structure in the inner disk
\citep[e.g.][]{1994AJ....108.1906H,1999AJ....118.1043H,1999A&A...349..619B,mcg15}
is strongly disfavoured.

We therefore are left to consider scenarios that could arise around a main-sequence or weak-line T Tauri star that has
dispersed its protoplanetary disk, but still hosts a gas-poor planetary system that may
include planets, asteroids, and comets.  The ``clumps'' of dust passing in front of the
star could perhaps lie within an optically thin asteroid belt analogue that is otherwise
undetected, or be more isolated objects such as remnants of a broken up comet.
As in the above scenarios, the typical minimum sizes of the dust 
grains are $\sim$$\mu$m \citep[e.g.][]{1993prpl.conf.1253B}, which are 
able to cause stellar variation by absorbing and scattering starlight at optical wavelengths.
Before considering such scenarios in more detail, we start with some scenario-independent
constraints that can be gleaned from the observations.

%%%%%%%%%%%%%%%
\section{Scenario-independent constraints}
\label{ss:indep}
\begin{figure}
  \begin{center}
    \includegraphics[trim=26mm 39mm 27mm 30mm, clip, width=84mm]
    {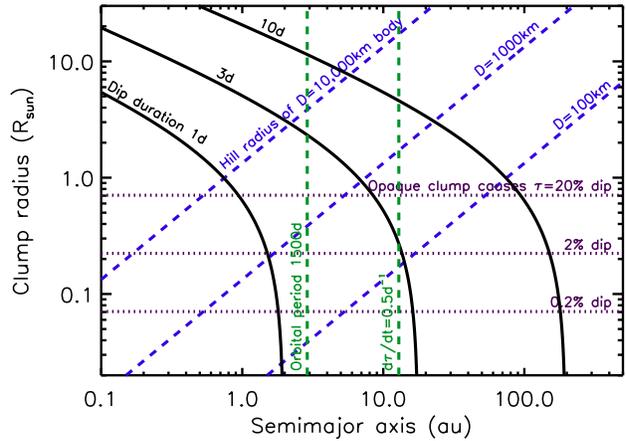}
  \vspace{-5cm}
    \caption{Size vs. semi-major axis parameter space for optically thick, spherical dust clumps on
      circular orbits around a star of $M_{\ast}=1.43$~\msun\ and $R_{\ast}=1.58$~\rsun.
      The solid lines represent dips of equal duration (as labelled).
      Dotted lines show minimum clump sizes for dips of different depths.
      Vertical dashed lines show where the orbital period is 1500 days, and where the
      light curve gradient for an optically thick ``knife edge'' could be as high as
      0.5 d$^{-1}$.
      Diagonal dashed lines show Hill radii of planetesimals of different sizes, assuming
      a density of 3~g~cm$^{-3}$.
      Combined, the period, gradient, and duration constraints in the circular
      orbit scenario suggests the clumps lie between $\sim 3 - 10$~AU, and have sizes similar
      to the star.}\label{fig:rcvsa}
  \end{center}
\end{figure}

To understand what could be the origin of the clumps it would help to know
where they are located in the system, how big they are, and how long they last.
To aid with this discussion, Figure~\ref{fig:rcvsa} shows some scenario-independent
constraints on the size and orbital distance of the clumps that are discussed further
below.
The only assumption for now is that the clumps are on circular 
orbits, but this assumption is relaxed later in Section~\ref{ss:comet}.
Some of the constraints also assume the clumps to be opaque, 
but again this assumption is relaxed later.

%%%%%%%%%%%%%%%
% begin figure rcvsav2
%%%%%%%%%%%%%%%
%%%%%%%%%%%%%%%

{\it Dip duration:}
The timescale $t_{\rm dip}$ for the transit of a clump of radius $s$ with transverse velocity
$v_{\rm t}$ across the equator of a star with radius $R_{\ast}$ is
$t_{\rm dip} = 2 \left( s + R_{\ast} \right) / v_{\rm t}$.
If the clump is on a circular orbit around a star of mass $M_{\ast}$ with semi-major axis $a$,
and is much less massive than the star, then
\begin{equation}\label{eq:clumpsize}
  s \approx 1.85 \, t_{\rm dip} \left( \frac{M_{\ast}}{a} \right)^{1/2} - R_{\ast},
\end{equation}
for $a$ is in units of AU, $M_{\ast}$ in \msun, $s$ and $R_{\ast}$ in \rsun, $t_{\rm dip}$ in days.
Thus, the several-day duration of the events for \thisstar\ suggests that the clumps are either
close-in and large compared to the star, or far-away from the star and small.
However, clumps that are too distant move too slowly across the stellar disk to explain the
observed duration regardless of their size;
e.g., a 3-day duration dip cannot arise from a clump beyond $\sim 15$~AU.

{\it Dip depth:}
A minimum clump size is set by the depth of the dimming events, which we
characterise as 1 minus the normalised flux, which we call $\tau$.
For example, even if the clump is completely opaque, the maximum dip depth is
${\rm max}(\tau)=(s/R_{\ast})^2$. 
The deepest $\tau=20$\% dimming event at D1500 thus implies that at least some clumps
are a sizeable fraction of the stellar size.
A dip caused by a fully optically thick symmetrical clump would also have a characteristic symmetrical shape
which does not resemble those observed (e.g., panel `c' in Figure~\ref{fig:kepler}), so this can be regarded as a strong
lower limit.
While there appear to be a range of event durations, the duration of the deepest events is
at most about 3 days.
The middle solid line in Figure~\ref{fig:rcvsa} (for $t_{\rm dip}=3$~d and a depth of $\tau=20$\%) therefore decreases
the outer limit on the clump locations mentioned above to closer to 8~AU.

{\it Light-curve gradient:}
A similar, but independently derived, outer constraint on the clump location can be set by
examining the gradients in the light-curve, which are at most half of the total stellar
flux per day (i.e. 0.5 d$^{-1}$ when the light curve is normalised to 1).
Orbiting material can change the light-curve most rapidly when it is optically thick and
passing the stellar equator \citep[i.e., the ``knife edge'' model of][]{2014MNRAS.441.2845V}.
The high rate of change in the \thisstar\ light curve translates to a lower limit on
the transverse velocity of the orbiting material of about 9~\kms, which corresponds
to an upper limit of 13~AU for material on circular orbits, although as 
discussed in Section~\ref{ss:comet}, this upper limit is closer to the star if the clump is optically thin.

{\it Non-periodicity:}
The lack of evidence for a clear periodicity in the dips in the observed light-curve excludes
orbital periods shorter than $\sim 1500$ days, which thus constrains the location
to lie beyond about 3~AU.
This constraint could be broken if the clumps disperse within a single orbit.  Likewise, 
if the two deep dipping events at D800 and D1500 are from the same 
orbiting body (or bodies), a period of 700 -- 800 days remains a possibility.

{\it Gravitational binding:}
To address the survival of the clumps, we note that in any scenario where the clumps
are not self-gravitating, they cannot be long-lived in the face of orbital shear
\citep[e.g.][]{ken05} and their internal velocity dispersion
\citep[e.g.][]{2012MNRAS.425..657J}.
Figure \ref{fig:rcvsa} therefore shows planetesimal sizes required to retain dust clouds
within their Hill sphere, $R_{\rm Hill} = a (M_{\rm pl}/[3M_{\ast}])^{1/3}$,
as one way of ensuring long-lived clumps.

Thus, under the assumption of circular orbits, the depth, duration and lack of
periodicity of the dimming events constrains their location to a region roughly
corresponding to that occupied by the giant planets in the Solar 
System (i.e., between the green dashed lines).
Clump sizes would thus be comparable to, but larger than, the 
star (i.e., above the uppermost horizontal dotted purple line), and they would 
have to have high, but not necessarily unity optical depth.
It might be possible to explain the clumps as dust bound to planetesimals larger
than around 1000~km, which means such planetesimals are not necessarily large
enough for direct transit detection (the lack of which could provide another constraint).

%%%%%%%%%%%%%%%%%%%%%%%%%%%%%%%%%%%%%%%%%%%%%%%%%%%%%%%%%%

\begin{figure}
  \begin{center}
    \includegraphics[width=84mm]{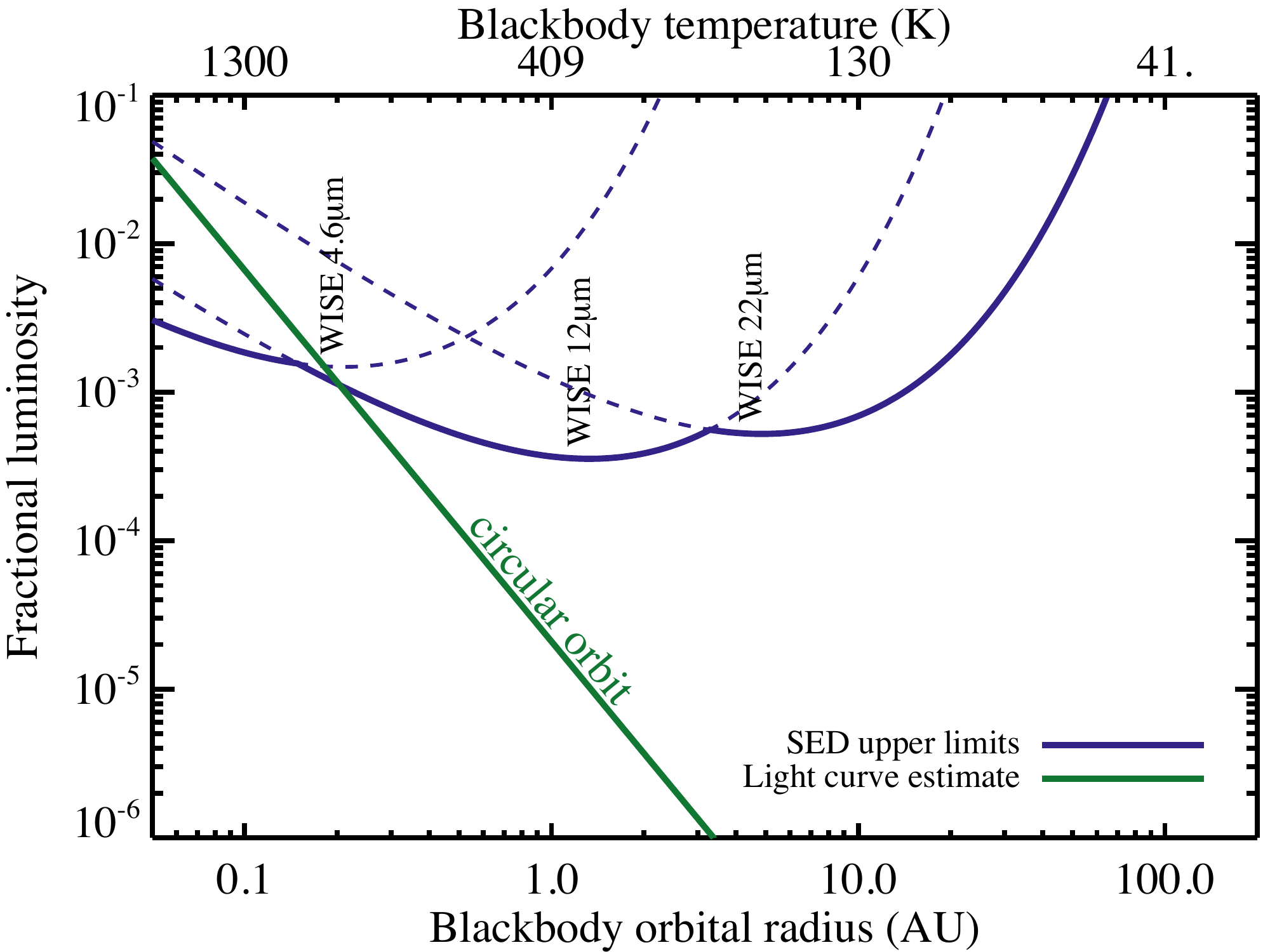}
    \caption{Fractional luminosity limits (blue lines) and an estimate of the system dust content from the light curve (green line). The dust level is constrained to lie below the blue line by the WISE photometry (4.6~$\mu$m, 12~$\mu$m, and 22~$\mu$m). The green line integrates the optical depth in the light curve assuming that clumps are similar in size to the star and on circular orbits. If the clumps lie beyond about 0.2~AU the IR non-detection of the dust is unsurprising, although many scenarios require more emission than that from dust seen to pass along our line-of-sight to the star.  Refer to Section~\ref{ss:indep} for details.}\label{fig:detlim}
  \end{center}
\end{figure}

%%%%%%%%%%%%%%%%%%%%%%%%%%%%%%%%%%%%%%%%%%%%%%%%%%%%%%%%%%
% inverted LC plot 
\begin{figure}
  \centering  
      \includegraphics[width=84mm]
      {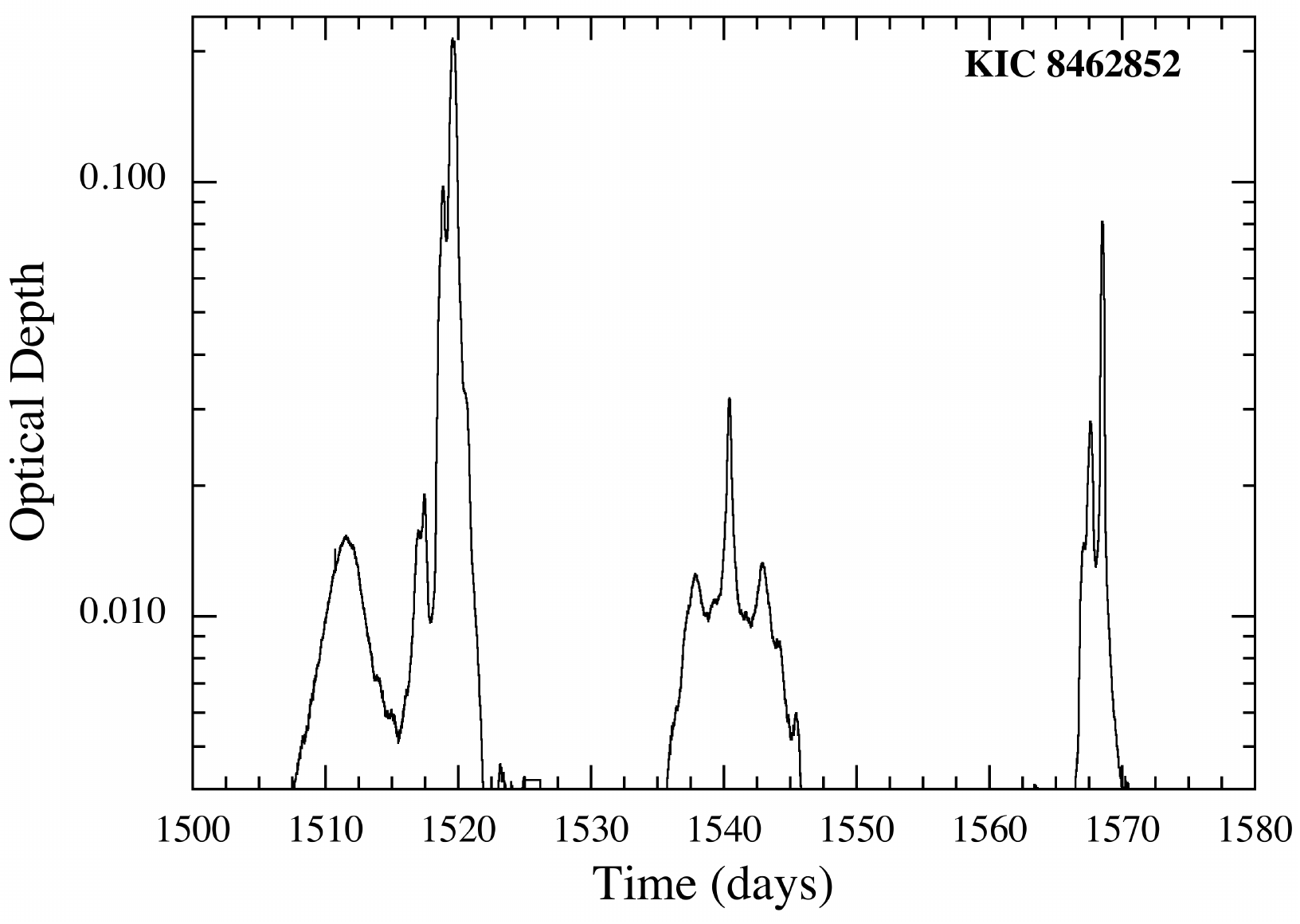}
    \caption{Inverted light curve for \thisstar\ portraying the blocking factors needed to reproduce the light curve as a function of time. Refer to Section~\ref{ss:indep} for details.}
    \label{fig:inverted_lc} 
  \end{figure} 
%%%%%%%%%%%%%%%%%%%%%%%%%%%%%%%%%%%%%%%%%%%%%%%%%%%%%%%%%%

{\it Infrared excess:}
Another constraint on the origin of the clumps comes from the lack of infrared emission
(Section~\ref{sec:sed}).
Assuming the clumps are larger than the star, the \kepler\ light curve provides 
blocking factors needed as a function of time, $\ln({\rm normalized~flux})$, 
where $\ln({\rm normalised~flux})\approx \tau$ for small $\tau$, 
as shown in Figure~\ref{fig:inverted_lc}.  
This optical depth and the assumption that the clump crosses the star at 
its orbital velocity allows conversion to optical depth as a function of distance along the clump.
The dimming events therefore allow an estimate of the minimum 
possible cross-sectional area $\sigma_{\rm tot}$ of dust in orbit around the star.  That is,
\begin{equation}\label{eq:dustint}
  \sigma_{\rm tot} = v_{\rm t} h \int \tau(t) dt,
\end{equation}
where the light-curve yields $\int \tau(t) dt \approx 0.86$ days,
$v_{\rm t}$ is the velocity of the clumps (assumed to be uniform at circular velocity
for a given semi-major axis), and $h$ the ``height'' of the clumps (i.e. their size along
the dimension perpendicular to their velocity).
The height of the clumps is assumed to be 2~R$_{\ast}$, though it could be higher if not all of the
clump crosses the stellar disk (e.g., this could be assumed to be $\pi s/2$ for large
spherical clumps passing directly across the star).  This calculation 
gives the minimum possible cross-sectional area as
\begin{equation}\label{eq:dustint2} 
\sigma_{\rm tot}=2.6 \times 10^{-4} a^{-1/2} {\rm AU}^2
\end{equation}
 where $a$ is in AU, the dependence on which arises from the velocity 
at which the clump crosses the star.

This cross-sectional area can then be converted to fractional luminosity at a given distance from
the star using $f = \sigma_{\rm tot} / (4 \pi a^2)$.  The blue lines in Figure~\ref{fig:detlim} show the limits on the dust fractional
luminosity $f=L_{\rm dust}/L_{\ast}$ derived from the SED (Section~\ref{sec:sed}).
These can be thought of as the maximum luminosity of blackbodies at a range of dust
temperatures (or stellocentric radii) that fit under the WISE photometry.
The dust estimate from Equation~\ref{eq:dustint} is shown as a green line, and the fact that it lies
below the blue line at all radii beyond 0.2~AU indicates that it is perhaps not particularly
surprising that no mid-IR excess was seen.

However, this dust area estimate is only a lower limit since it only includes the dust which passed
in front of the star during the lifetime of the \kepler\ mission.
The true area would be larger if there are more clumps further along the orbit which
have yet to pass in front of the star, and could also be larger if the dips do not capture
all of the cross-sectional area in their clumps.
Furthermore, for some specific scenarios discussed in the following sections, the presence of clumps that pass in front of the star
requires the existence of other clumps that do not pass along our line-of-sight.
The lack of infrared emission thus places constraints on how many such clumps there are in
the system.  For example, Figure~\ref{fig:detlim} shows that for clumps at a few AU 
the cross-sectional area can only be increased
by 3 orders of magnitude before it is detectable by WISE.
The calculation is further complicated should the clumps be considered to be short-lived,
or on non-circular orbits.

{\it Mass estimates:}
The minimum possible cross-sectional area required to cause the observed dips, $\sigma_{\rm tot}$ (Equation~\ref{eq:dustint2}), can also be used to determine a minimum possible dust mass, $m_{\rm tot}$. If the dust all has the same diameter $D$ and density $\rho$ then $m_{\rm tot}/\sigma_{\rm tot}=2\rho D/3$, resulting in a total mass of $6.7 \times 10^{18}$~g for 1~$\mu$m diameter dust of density 3~g~cm$^{-3}$ orbiting at 3~AU (and scaling as $a^{-1/2} \rho D$ for different assumptions). If all of this mass were put in a single body of the same density $\rho$, this would have a diameter of 16~km. This illustrates that the minimum mass of the parent body required to cause this phenomenon is approaching the mass of comet Hale-Bopp. However, this calculation has 
a few caveats. For one, the value derived for $\sigma_{\rm tot}$ is an absolute minimum given that it only accounts for the material which passed in front of the star during the observations. It also does not account for the possibility that the dust in the clump has a range of sizes. For example, for dust with a power law size distribution with index of 3.5  \citep{1969JGR....74.2531D} extending from $D_{\rm min}$ to $D_{\rm max}$, the ratio of mass to cross-sectional area scales $\propto \rho \sqrt{D_{\rm min}D_{\rm max}}$. Thus, this estimate would be 100 times larger than that derived above if the size distribution extended from 1~$\mu$m up to 1~cm. 

Given these basic constraints we now consider several scenarios that may explain the
observations.
The first two are related to collisions within an asteroid belt (Section~\ref{ss:ab})
or unstable planetary system (Section~\ref{ss:gi}), the third considers dust that orbits
within the Hill spheres of large planetesimals which may reside in an asteroid belt
but are not required to collide (Section~\ref{ss:hill}),
and the fourth is that the dips are the passage of a series of fragments from a broken-up
comet or asteroid on a highly elliptical orbit (Section~\ref{ss:comet}).

%%%%%%%%%%%%%%%
\section{Specific Occultation Scenarios}
\label{ss:specific_occ_scen}
\subsection{Aftermath of catastrophic collisions in asteroid belt}
\label{ss:ab}
One possibility is that the dimming events are caused by dust thrown off in collisions
between planetesimals in an otherwise unseen asteroid belt analogue (e.g., \citealt{2002MNRAS.334..589W,2014MNRAS.439..488Z}).
The dust clouds created in these destructive collisions expand at roughly the planetesimals' escape
velocity from the colliding bodies, eventually spreading and shearing out to form a
smooth dust component in which the clumps reside.
Such a scenario is a promising explanation for the star RZ~Psc \citep{2013A&A...553L...1D},
though in that case evidence that the underlying asteroid belt exists is given by a strong
IR excess.

There are several problems with this scenario as applied to \thisstar\ however. 
Probably the most fundamental of these is the absence of 
an IR excess from the smooth component.
This is because for every clump we see, remembering that these were inferred to be
slightly larger than the star, there should be many more that have spread out.
The infrared emission from the dispersed clumps would likely sum up to a detectable
level, even before counting dust produced in non-dip forming events.
Moreover we should see dips from the clumps in the middle of being dispersed (i.e., dips
with longer duration albeit lower optical depth), as well as dips with a continuum of depths and
durations from the many different scales of planetesimal impacts that would occur.
The clustering of dips at D1500 also points to these events being correlated which is hard to reconcile
with this scenario, though the planetesimals in the belts could be shepherded by planets
into confined azimuthal regions (e.g., \citealt{2003ApJ...598.1321W,2013ApJ...768...45N}).

%%%%%%%%%%%%%%%
\subsection{Aftermath of giant impact in planetary system}
\label{ss:gi}
A possible way around the issues in Section~\ref{ss:ab} is to invoke dust thrown off in a single
collision, perhaps analogous to the Earth-Moon system forming event \citep{2012MNRAS.425..657J}.
In this case there need not be an underlying asteroid belt, as the collision could be
between planets whose orbits recently became unstable, or between growing planetary embryos.
Such events are expected to result in strong IR excesses
\citep[e.g.][]{2012MNRAS.425..657J,2015arXiv150800977G}, and are indeed seen in 
systems such as HD\,172555 where giant impacts are the favored explanation \citep{2009ApJ...701.2019L}. 
In this scenario, the putative collision would need to have occurred between the WISE 
observation taken in \kepler\ Q5 and the first large dip at D800.
The dip at D1500 is then interpreted as the same material seen one orbit later,
with the $\sim 750$ day period implying an orbit at $\sim 1.6$\,AU.
The difference in the dip structure from D800 to D1500 could arise because the clump(s) created in the original
impact are expanding and shearing out.
This scenario therefore predicts that \thisstar\ may now have a large mid-IR excess,
but the most recent IR observations taken in 2015 January with Spitzer IRAC
show no significant excess for \thisstar\ \citep{2015ApJ...814L..15M}. 
However, non-detection of an excess would not necessarily rule this scenario out, as the dust levels
derived in Section~\ref{ss:indep} (which account for the dust seen passing in front of the
star) were shown to be consistent with a non-detection.
A more robust prediction is that future dimming events should occur
roughly every 750 days, with one in 2015 April and another in 2017 May.

Two new issues arise with this scenario however.
Firstly, if the period of the orbiting material is a few years, what is the origin of the
two small $0.5$\% dips seen in the first few hundred days (D140 and D260; Table~\ref{tab:Dip-Table}), and why did they not repeat 750 days later?
It is a concern that these could require the existence of an outer planetesimal
belt, which may contradict the lack of infrared emission to this star.
Perhaps more problematic is the probability that this
star (of unknown age) should suffer such an event that occurs within a few-year window
between the WISE observation and the end of the prime \kepler\ mission, and that the
geometry of the system is such that material orbiting at $\sim$1.6~AU lies almost exactly
between us and the star.
Taking this few year window, the main sequence lifetime, and an optimistic estimate for the
scale height of giant impact debris, and the number of \kepler\ stars observed, this suggests
that every star would have to undergo $10^4$ such impacts throughout its lifetime for us to
be likely to witness one in the \kepler\ field.
Thus, while this scenario is attractive because it is
predictive, the periodicity argument may be inconsistent, and the probability of
witnessing such an event may be very low (though of course difficult to estimate).

%%%%%%%%%%%%%%%
\subsection{Dust-enshrouded planetesimals}
\label{ss:hill}
Scenarios in which the clumps can be long-lived are attractive because they suffer less
from being improbable.
Thus, one possibility is that the clumps are held together because they are in fact
themselves orbiting within the Hill sphere of large planetesimals.
They can therefore be thought of as planetesimals enshrouded by near-spherical swarms of
irregular satellites, which are themselves colliding to produce the observed dust.
This scenario is therefore analogous to that suggested for the enigmatic exoplanet Fomalhaut b
\citep{2008Sci...322.1345K,2011MNRAS.412.2137K}, which borrows from the irregular
satellites seen in the Solar System
\citep[e.g.][]{2007ARA&A..45..261J,2010AJ....139..994B}.
This scenario suffers from several problems.
First, the observed dips already require multiple large planetesimals.
Unless these all orbit within the same plane to a high degree (i.e., to within a few
stellar radii), there must be many more large planetesimals which never (or have yet to)
pass in front of the star.
Debris disks with low levels of stirring are theoretically possible \citep{2010MNRAS.401..867H,2013ApJ...772...32K}.
However, these low stirring levels require the absence of large planetesimals which through mutual interactions
would stir the relative velocities to their escape speeds.
This is in addition to the problem of filling the Hill sphere of such planetesimals almost
completely with dust.
This may be reasonable if the planetesimals are embedded in a belt of debris.
However, that would incur the problem of the lack of infrared excess.
The question also remains why the D1500 events are so clustered, and why 
there are several deep dimming events and no intermediate ones.
A population of planetesimals should have a variety of inclinations with respect to
our line of sight, so they should pass in front of the star at a range of impact
parameters and cause a range of dip depths.

A related scenario is that the planetesimals are surrounded by large ring systems,
similar to that invoked to explain the $\sim$50 day dimming event seen for 1SWASP
J140747.93-394542.6 \citep[normally called
``J1407'',][]{mam12,2014MNRAS.441.2845V,2015ApJ...800..126K}.
In that case however, a single relatively time-symmetric dimming event was seen,
whereas \thisstar\ has multiple asymmetric events.
Thus, a single ringed planet(esimal) would not reproduce the observed light-curve,
and a scenario with multiple ringed-planetesimals would be essentially the same as
the irregular satellite scenario above.

%%%%%%%%%%%%%%%
\subsection{A family of objects on a comet-like trajectory}
\label{ss:comet}

One of the scenario independent constraints considered in Section~\ref{ss:indep} was 
the presence of light-curve gradients as large as 0.5\,d$^{-1}$, which results in an upper 
limit of 13\,AU for the clumps' semi-major axis 
assuming optically thick clumps (Figure~\ref{fig:rcvsa}).
However, the star is never completely occulted, so this estimate should be corrected for the
optical depth of the clump $\tau$.
That is, the steepness of the gradient is diluted either by flux transmitted through a large
optically thin clump (or by unocculted parts of the star for an optically thick small clump).
Assuming $\tau=0.2$ the velocity estimate given by the gradients is then 5 times
higher than assumed in Section~\ref{ss:indep};
this would predict a more realistic minimum transverse velocity of $\sim$50~\kms\ to 
cause the observed gradient, which for a circular orbit yields a maximum semimajor axis of $a=0.5$~AU.
While this estimate is uncertain, for example because of the unknown optical depth structure of
the different clumps, this highlights the possibility that the material may be moving so
fast that the velocity for a circular orbit is inconsistent with the non-repetition of the events.

\begin{figure}
  \begin{center}
          \includegraphics[trim=26mm 39mm 27mm 30mm, clip, width=84mm]
      {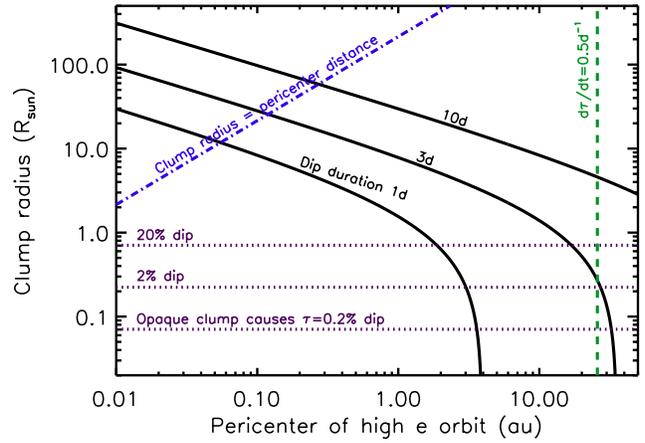}
        \vspace{-5cm}
    \caption{Size vs. pericenter parameter space for high eccentricity comet-like orbits. 
      Dotted lines show lower limits on the clump sizes from the dip depths. The
      dashed line is the outer limit set by the light curve gradient, 
      noting that this limit decreases with decreasing optical depth, e.g., the limit would be at a pericentre that is 25 times smaller than that plotted if the clumps have optical depth of 0.2 (line not shown in figure). 
      The dot-dashed line
      is where the clump radius equals the pericenter distance, though the clumps could exist
      above here if they are elongated along the orbital direction. The solid lines are 
      of constant dip duration.}\label{fig:rcvsq}
  \end{center}
\end{figure}

One solution to this problem is that the orbits need not be circular.
That is, we could be seeing material close to the pericenter of a highly eccentric orbit,
reminiscent of comets seen in the inner Solar System at pericenter \citep{1967AJ.....72.1170M,1984Icar...58...81S}.
Comets around other stars have also been detected, the first of these being found around Beta Pictoris
 \citep{1984Sci...226.1421S,1989A&A...215L...5L,1990A&A...236..202B}.  We therefore 
 envision a scenario in which the dimming events are caused by the passage of a
series of chunks of a broken-up planetesimal on a comet-like orbit. That planetesimal 
may have been analogous to what we refer to in the Solar System as a `comet', in 
which case it could be volatile-rich and may have broken up as a result of thermal 
processes. However, it may alternatively have closer analogy with Solar System 
asteroids in having a more refractory composition, which might require non-thermal 
processes such as tidal disruption to break it up. The disruption mechanism and 
composition of the planetesimal are not defined for this scenario, just its orbit 
which is comet-like, and so we refer to it here-on as a ``comet-like'' without bias to their 
origin or physical make-up.  Regardless of its disruption process, the resulting chunks would have to have since
spread around the orbit, and may be continuing to fragment to cause the erratic nature of the observed dips.

To assess this scenario, Figure~\ref{fig:rcvsq} revisits the clump - orbit parameter space
of Figure~\ref{fig:rcvsa} (discussed in Section~\ref{ss:indep}), but now uses the pericenter of the
clump's orbit instead of its semi-major axis.
The orbits are assumed to be highly eccentric ($e \approx 1$), with the dips arising from material
close to pericenter, so that their orbital velocity is roughly $\sqrt{2}$ times the circular Keplerian velocity
at that distance.
The limits from the dip depths and light-curve gradient are again shown, as
are lines of constant dip duration.
The planetesimal Hill radius lines are not shown, because they are not applicable to the cometary
scenario considered here, though these would be slightly modified versions of those in
Figure~\ref{fig:rcvsa} (see eq. B5 of \citealt{2014MNRAS.443.2541P}).
In general, the main change compared with Figure~\ref{fig:rcvsa} is that the higher orbital
velocity relaxes the constraints on how far out the clumps can be orbiting.
However, as mentioned above, if the clumps are optically thin (as opposed to optically thick as assumed 
in Figure~\ref{fig:rcvsq}) the constraint from the light curve gradient may be more stringent. 
For example, decreasing the optical depth to 0.2 would result in a transverse velocity 
of 50~\kms\ (see above), thereby moving the light curve gradient constraint on the upper limit from 26\,AU closer to 1\,AU.

The proximity of the comet-like clump to the star when causing the dip does not present a problem for 
this scenario, as it did when the clump was on a circular orbit.  This is because the pericenter distance does 
not necessarily bear any relation to the period with which the comet-like fragments return to 
pass in front of the star. That period is set by the semimajor axis which has the same 
constraint as shown on Figure~\ref{fig:rcvsa}, and there is no such constraint 
on the pericenter in Figure~\ref{fig:rcvsq}.  Thus the point of note from 
Figure~\ref{fig:rcvsq} is that the pericenter could be significantly within 1\,AU.
Closer pericenters are favored both because this geometry results in a higher probability of
the clumps occulting the star along our line-of-sight, and because of the greater opportunities
for fragmentation of the bodies.
The temperatures of comets (i.e., with volatiles) at such close proximity to the star ($>410$~K) would render them
susceptible to thermal stresses.
The existence of multiple super-Earth planets orbiting $<1$\,AU from many main sequence stars
also points to the possibility that the body could have been tidally disrupted in a close
encounter with one such planet.
It is even possible that the body came close enough to the star for tidal disruption in
the absence of other considerations;
e.g., a comet similar to Halley's comet would fall apart by tidal forces on approach to 
within 3--7 stellar radii (0.02 -- 0.05~AU). By contrast, a rocky body would require
a closer encounter to tidally disrupt.

For close pericenters it is important to point out that while the constraint is discussed in
terms of the clump's radius, the clump can not in fact be spherical at that size.
Figure~\ref{fig:rcvsq} shows a blue dot-dashed line where the ``clump radius'' is the same as
the pericenter distance.
At such proximity, the clump could not be elongated in the radial direction, but could
only be elongated azimuthally along the orbit.
In fact, this mostly linear clump structure is the correct way to visualize debris from 
the breakup of a comet or planetesimal.  The small velocity kicks 
(from fragmentation or tidal disruption) would cause a small dispersion 
in semimajor axis for material in the clump, and the resulting differential 
orbital motion causes the material to spread around the orbit. These small 
kicks do not significantly change the periastron distance or the orbital inclination angle.

This scenario is attractive, because comets are
known in the Solar System to have highly eccentric orbits and disrupt for various reasons
near pericenter, and infalling comets are the most robust
explanation for the falling
evaporating body (FEB) phenomenon seen around many nearby A-type stars
\citep[e.g.][]{1985ApJ...291L...1K,1990A&A...236..202B,wel13, 2014Natur.514..462K}.
Also, since fragments of the comet family would all have very similar orbits, this
mitigates the problem noted in Section~\ref{ss:ab} that the detection of multiple transits may require
orders of magnitude more clumps to be present in the system.
Instead, the observed clumps may be essentially in a single orbit which is that of the progenitor, 
and that orbit happens to be preferentially aligned for its transit detection.
That is, it is not excluded that we have observed all the clumps present in the system.
While a quick look at Figure~\ref{fig:detlim} suggests that the lack of infrared excess
might still be problematic for the closest pericenters (noting that $\sigma_{\rm tot}$ also 
needs to be increased by $\sqrt{2}$ due to the higher transverse velocity at pericenter in 
Equation~\ref{eq:dustint2}), in fact that is not necessarily the case.
Rather, in that figure we assumed that the clumps were present at the given distance at all times,
whereas the clumps in the comet-like group scenario were at much larger separation from the
star at the time of the WISE observations.  The total mass of the fragmented body 
was considered in Section~\ref{ss:indep}, but since the clumps can be closer to the star 
in this scenario, and are moving faster than circular Keplerian velocity, a better minimum 
mass estimate for clumps seen at a pericenter of 0.1\,AU is $\sim 3 \times 10^{19}$~g. 
Again, the size distribution and any material not contributing to the observed dips will 
increase this minimum mass, perhaps by a factor of 100, leading to a more realistic 
parent body mass of $3 \times 10^{21}$~g , consistent with a rocky body $\sim 100$km in diameter.

It remains to be shown that this model can explain the more detailed structure of
the light-curves.
Some potential positives are that the clustered nature of the dips could be explained
by subsequent fragmentation of a large fragment from an earlier break-up.
The smaller dips could also potentially be explained by smaller fragments which may
also be expected to receive larger kicks during fragmentation.
However, the structure of individual clumps may be problematic. 
For example, a fairly generic prediction of transits of
comet-like bodies may be that their light-curves show signs of their tails.
The light-curve expected for a typical event then has a relatively fast ingress as the head of
the comet passes in front of the star, but a slower egress as the tail passes
\citep[e.g.][]{1999A&A...343..916L,rap12}.
However, the D800 event shows the opposite (see panel `c' in Figure~\ref{fig:kepler}).
Possible resolutions of this issue are that the D800 comet fragment received a large
kick with an orientation that sheared it out in such a way to form a ``forward tail''.
Such forward comet tails produced by the fragments being kicked toward the 
star have been studied in the literature, but require the grains in the tail to be large enough to overcome 
the effects of radiation pressure \citep{san15}.
Alternatively, this event could be comprised of two dips superimposed to have the appearance
of a forward tail.
While several issues remain to be explored, of the scenarios considered we conclude
that a cometary-like group of bodies seems most consistent with the data at hand.

%%%%%%%%%%%%%%%%%%%%%%%%%%%%%%%%%%%%%%%%%%%%%%%%%%%%%%%%%%%%%%%%
\section{Summary and Conclusions}\label{sec:discussion}   

In this paper, we have shown that  \thisstar\ is a unique source in the \kepler\ field.  This otherwise seemingly normal F star undergoes erratic and completely unpredictable dips in flux ranging from $\lesssim 1\%$ to more than 20\%.  Most of the approximately 7 dips observed before D1500 have fairly smooth, but unexplained, dip profiles that are each several days long.  The D1500 sequence lasts continuously for at least 80 days, but the majority of that time is spent with the flux depressed by less than $\sim$2\%. 

We have conducted numerous follow-up investigations of the star and its environment, including spectroscopy, adaptive optics imaging, construction of a spectral energy distribution, generation of a Fourier transform and a sonogram using the \kepler\ time series, and examination of ground-based photometry.  Our analysis characterizes the object as both remarkable (e.g., the ``dipping'' events in the \kepler\ light curve) and unremarkable (ground-based data reveal no deviation from a normal F-type star) at the same time.  

An extensive set of scenarios has been presented to explain the occurrence of the dips, most of which are unsuccessful in explaining the observations in their entirety.  Of these, the scenarios invoking intrinsic variability, such as the Be star framework, were deemed unlikely, but they are not entirely ruled out as a plausible option to explain the dips.  However, we pointed out that the relatively low $T_{\rm eff}$ and lack of H$\alpha$ emission and IR excess in KIC 8462852 are not suggestive of Be-star activity.

A broad range of scenarios for the dipping behavior that involve occultation by circumstellar dust clumps was considered.  Among these, we find that the break-up of one or more massive exocomets (or planetesimals on comet-like orbits) provides the most compelling explanation consistent with the data in hand. The required mass of the original body may have been in excess of $3 \times 10^{21}$ grams (only $\sim$0.3\% the mass of Ceres, and perhaps $\sim$100 km in diameter).  

We can envision a scenario in which a barrage of bodies, such as described above, could be triggered by the passage of a field star through the system.  And, in fact, as discussed in Sect.~\ref{sec:imaging}, there is a small star nearby ($\sim 1000$ AU; Section~\ref{sec:imaging}) which, if moving near to \thisstar, but not bound to it, could trigger such a barrage into the vicinity of the host star.  On the other hand, if the companion star is bound, it could be pumping up comet eccentricities through the Kozai mechanism. Measuring the motion/orbit of the companion star with respect to \thisstar\ would be telling as to whether or not they are physically associated, and we could then be better able to make assessments about the timescale and repeatability of comet showers based on bound or unbound star-comet perturbing models. 

Continuing observations of \thisstar\ should aid in unraveling the peculiar dips in its light curve. First and foremost, long-term photometric monitoring is imperative in order to catch future dipping events. It will be helpful to know whether such observations reveal continued, possibly periodic dips, or no further dips.  If the dips continue, it will be important to search for a clear periodicity, and to look for changes in depth or shape. To completely solidify the hypothesis that the dips are due to dust, observations should study the wavelength dependence of the obscuration soon after a new dip is discovered.   In the case of a family of giant comet-like bodies there
presumably should be at least a few events similar to those seen with \kepler\ over the next decade.  However, if the comet-like objects actually populate a very long eccentric orbit (i.e., that of the original planetesimal), the material may be spread out around that orbit, and future dippings events could continue to appear over hundreds of years. 

Several of the proposed scenarios are ruled out by the lack of observed IR excess (Section~\ref{sec:sed}), but the comet/planetesimal fragments scenario has the least stringent IR constraints.   In the comet scenario, the level of emission could vary quite rapidly in the near-IR as clumps pass through pericenter (close to the time they are transiting) and are shedding new material. 
If the system is currently in the aftermath of a giant impact, there could be a semi-steady increase in IR flux over years/decades.  The WISE observations were made in Q5, and assuming that an impact occurred in Q8 (D800, Section~\ref{ss:gi}), detecting the IR emission from such an impact is still a possibility in the future.  The only Spitzer IRAC observation of \thisstar, taken in January 2015, showed a marginal, but below 3-$\sigma$, excess at $4.5 \mu$m, disfavoring the impact scenario \citep{2015ApJ...814L..15M}.  Continued monitoring in the IR will allow us to firmly distinguish between the giant-impact and cometary-group scenarios.

In summary, it will require some observational skill and patience to find the next dipping event from this object using ground-based observations.  As we pointed out, the source spent a rather small fraction of its time during the 4-year {\em Kepler} mission with dips of greater than 2\%.  Nonetheless, the key to unraveling the mysterious dips will require such observations.

%%%%%%%%%%%%%%%%%%%%%%%%%%%%%%%%%%%%%%%%%%%%%%%%%%%%%%%%%%%%%%%%
%%%%%%%%%%%%%%%%%%%%%%%%%%%%%%%%%%%%%%%%%%%%%%%%%%%%%%%%%%%%%%%%
\section*{Acknowledgments}

We thank Jason Wright and Jason Curtis for fruitful discussions on the object. We further acknowledge Planet Hunter user ``Exoplanet1'' for their contributions to the discussion of this object. We are grateful to Sherry Guo and Bhaskar Balaji for running an automated search through the Kepler set to find other similar dippers. We acknowledge Mike Jura for very insightful comments about the required mass of the body that is the origin of the obscuring material. We thank Josh Carter for pointing out the possible 48-day periodicity.  We appreciate the efforts of Jeff Coughlin, Jon Jenkins, and Jeffrey Smith for taking a careful look at the raw \kepler\ photometry to decided if it was all good, i.e., not artifacts. We thank Mark Everett and Lea Hirsch for making the DSSI observations.  We thank Huan Meng, Massimo Marengo, and Casey Lisse, for insightful comments related to the IR excess. We are grateful for thoughtful discussions with members of the \kepler\ Eclipsing Binary Working Group, and attendees of the K2 SciCon 2015. Last but not least, we are grateful for the anonymous referee's comments to help improve the paper.     

TSB acknowledges support provided through NASA grant ADAP12-0172 and ADAP14-0245.  MCW and GMK acknowledge the support of the European Union through ERC grant number 279973.  The authors acknowledge support from the Hungarian Research Grants OTKA K-109276, OTKA K-113117, the Lend\"ulet-2009 and Lend\"ulet-2012 Program (LP2012-31) of the Hungarian Academy of Sciences, the Hungarian National Research, Development and Innovation Office -- NKFIH K-115709, and the ESA PECS Contract No. 4000110889/14/NL/NDe. This work was supported by the Momentum grant of the MTA CSFK Lend\"ulet Disk Research Group.  GH acknowledges support by the Polish NCN grant 2011/01/B/ST9/05448.  Based on observations made with the Nordic Optical Telescope, operated by the Nordic Optical Telescope Scientific Association at the Observatorio del Roque de los Muchachos, La Palma, Spain, of the Instituto de Astrofisica de Canarias. This research made use of The Digital Access to a Sky Century at Harvard (DASCH) project, which is grateful for partial support from NSF grants AST-0407380, AST-0909073, and AST-1313370.  The research leading to these results has received funding from the  European Community's Seventh Framework Programme (FP7/2007-2013) under grant agreements no. 269194 (IRSES/ASK) and no. 312844  (SPACEINN).  We thank Scott Dahm, Julie Rivera, and the Keck Observatory staff for their assistance with these observations. This research was supported in part by NSF grant AST-0909222 awarded to M.\ Liu. The authors wish to recognize and acknowledge the very significant cultural role and reverence that the summit of Mauna Kea has always had within the indigenous Hawaiian community. We are most fortunate to have the opportunity to conduct observations from this mountain.  KS gratefully acknowledges support from Swiss National Science Foundation Grant PP00P2\_138979/1. HJD and DN acknowledge support by grant AYA2012-39346-C02-02 of the Spanish Secretary of State for R\&D\&i (MINECO). This paper makes use of data from the first public release of the WASP data \citep{but10} as provided by the WASP consortium and services at the NASA Exoplanet Archive, which is operated by the California Institute of Technology, under contract with the National Aeronautics and Space Administration under the Exoplanet Exploration Program. This publication makes use of data products from the Wide-field Infrared Survey Explorer, which is a joint project of the University of California, Los Angeles, and the Jet Propulsion Laboratory/California Institute of Technology, and NEOWISE, which is a project of the Jet Propulsion Laboratory/California Institute of Technology. WISE and NEOWISE are funded by the National Aeronautics and Space Administration. This research made use of the SIMBAD and VIZIER Astronomical Databases, operated at CDS, Strasbourg, France (http://cdsweb.u-strasbg.fr/), and of NASA's Astrophysics Data System.

%%%%%%%%%%%%%%%%%%%%%%%%%%%%%%%%%%%%%%%%%%%%%%%%%%%%%%%%%%%%

% Bibliography %

%\clearpage

%\bibliographystyle{apj}            % Please learn to use the
\bibliographystyle{mn2e}           % Please learn to use the
%\bibliographystyle{asp2010}         % formatting of Latex's Bibtex. It
                                    % will make your life easier.
% apj.bst should be in this directory as well as apj-jour.bib and reference paper.bib

\bibliography{mn-jour,paper}      % "paper.bib" contains all my
%\bibliography{paper}      % "paper.bib" contains all my
                                    % references. "apj-jour.bib"
                                    % contains abbreviations of
                                    % journals.

%%%%%%%%%%%%%%%%%%%%%%%%%%%%%%%%%%%%%%%%%%%%%%%%%%%%%%%%%%%%
%%%%%%%%%%%%%%%%%%%%%%%%%%%%%%%%%%%%%%%%%%%%%%%%%%%%%%%%%%%%

\end{document}